\documentclass[preprint,12pt]{elsarticle}
\usepackage{graphicx,psfrag,epsf}
\usepackage{amsfonts}
\usepackage{mathtools}
\usepackage{amssymb}
\usepackage{amsmath}
\usepackage{longtable}
\usepackage{bigints}
\usepackage{soul}
\usepackage{tikz}
\usetikzlibrary{arrows.meta}
\tikzset{>={Stealth[length=3mm]}} % better arrow heads
\usepackage{pgfplots}
\pgfplotsset{compat=1.16}
\usepackage{color}
\usepackage{arydshln}
\usepackage{hyperref}
\usepackage[capitalise]{cleveref}
\usepackage{url}
\usepackage{doi}
\usepackage{multicol}
\usepackage{multirow}
\usepackage{subcaption}

\newcommand{\x}{\boldsymbol{X}}

\newcommand{\normal}{\mathcal{N}}
\newcommand{\lexp}{\underline{\mathbb{E}}}
\newcommand{\uexp}{\overline{\mathbb{E}}}

\newcommand{\added}[1]{\textcolor{blue}{#1}}

\begin{document}

\begin{frontmatter}

%%% MT what about using "Robust Bayesian causal estimation" (name for our method) in the title, e.g. "Robust Bayesian causal estimation for causal inference in medical diagnosis"
\title{\added{Robust Bayesian causal estimation for causal inference in medical diagnosis}}
\author[1]{Tathagata Basu\corref{cor1}}
\ead{tathagatabasumaths@gmail.com}
\cortext[cor1]{Corresponding author}
\author[2]{Matthias C.~M.~Troffaes}
\ead{matthias.troffaes@durham.ac.uk}
\affiliation[1]{organization={Civil and Environmental Engineering, University of Strathclyde},
            addressline={16 Richmond St}, 
            city={Glasgow},
            postcode={G1 1XQ}, 
            %state={},
            country={United Kingdom}}
\affiliation[2]{organization={Department of Mathematical Sciences, Durham University},
            addressline={South Road}, 
            city={Durham},
            postcode={DH13LE}, 
            %state={},
            country={United Kingdom}}

\begin{abstract}
\added{Causal effect estimation is a critical task in statistical learning} that aims to find the causal effect on subjects
by identifying causal links between
a number of predictor (or, explanatory) variables and the outcome of a treatment. \added{In a regressional
framework, we assign a treatment and outcome model to
estimate the average causal effect.} Additionally, for high dimensional \added{regression problems, variable selection methods are also used} to find a subset of 
predictor variables \added{that} maximises the predictive performance of the \added{underlying model for better estimation of the causal effect.}  In this paper, we \added{propose a different approach. We focus on the variable selection aspects of
high dimensional causal estimation problem. We suggest a cautious
Bayesian group LASSO (least absolute shrinkage and selection operator) framework for variable selection using prior sensitivity analysis. We  argue that in some cases, abstaining from selecting (or, rejecting) a predictor is beneficial and we should gather more information to obtain a more decisive result. We also show that for problems with very limited information, expert elicited variable selection can give us a more stable causal effect estimation as it avoids overfitting. Lastly, we carry a comparative study with
synthetic dataset and show the applicability of our method in
real-life situations.}

\end{abstract}


\iffalse
\begin{highlights}
\item \added{A robust variable selection method for high dimensional causal estimation} 

\item \added{Prior sensitivity analysis of spike and slab group-LASSO.} 

\item \added{We show importance of elicitation in variable
selection problem}

\item \added{We show empirical behaviour of our method and
its usefulness in medical diagnosis.}

\end{highlights}
\fi
\end{frontmatter}
\section{Introduction}\label{sec:intro}

Causal inference using observational data is important in
many fields, including epidemiology, social science, economics, and many more.
Causal inference concerns estimating the causal
effect of predictor variables on an outcome variable,
as well as identifying which predictors are causally linked with the outcome.
Ideally,
randomised trials are the most efficient way to perform this task.
However, this is not always practical due to, for instance, ethical 
concerns, design cost, population size, to name a few. This
leaves us with observational studies
where data is collected though surveys or record keeping.

\added{%
Unfortunately, without fully controlled randomised trials and full
knowledge of confounders, it is well understood that
statistical models are unable to infer causality, as correlation
does not imply causation especially in the presence of confounders.
Still, it is highly desirable to try to adjust for confounding in
our statistical models to the best of our ability. This is termed
(perhaps somewhat unfortunately) \emph{causal inference} in the
literature
\citep{rubin1978,rosenbaum83,Robins1986ANA,winship99,stuart10,Zigler2014,wang2015,koch2018,Hahn2018,koch2020}. 
This is also the approach that we will follow here, under the
disclaimer that whether actual causality can be inferred
remains a subject of interpretation and conjecture specific to the
situation being studied. In this regard, we also refer to \citep{imbiens_2022}
where a detailed discussion on different interpretations of `causality'
in statistics and econometrics can be found 
}

\begin{figure}[h]
\begin{center}
\begin{tikzpicture}[node distance=2cm, auto]
  \node (confounder) at (0,0) [circle,draw,align=center] {biomarker \\(confounder)};
  \node (predictor) at (4,2) [circle,draw,align=center] {treatment \\ decision \\ (predictor)};
  \node (outcome) at (4,-2) [circle,draw,align=center] {treatment \\ outcome};
  \draw[->] (confounder) to (predictor);
  \draw[->] (confounder) to (outcome);
  \draw[->,dashed] (predictor) to (outcome);
\end{tikzpicture}
\end{center}
\caption{A biomarker influencing both the treatment decision and the treatment outcome, thereby acting as confounder. Solid arrows indicate causation, whilst the dashed arrow indicates correlation without causation.}
\label{fig:confounding}
\end{figure}
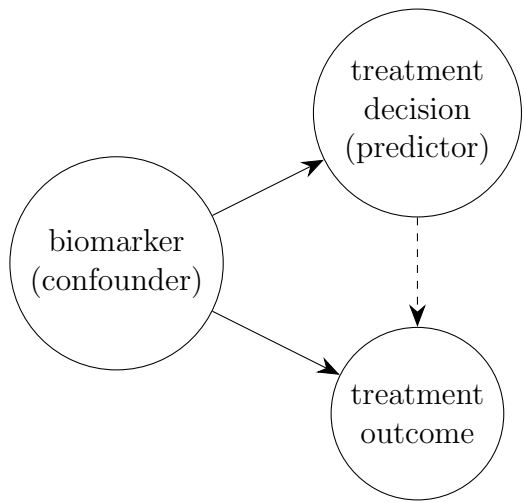
A confounder is any variable which is causally linked with both a predictor and the outcome, giving the false impression that that predictor causes the outcome (see \cref{fig:confounding}).
Confounding happens commonly in observational treatment studies because many predictors are often causally linked with the treatment decision (which is also a predictor), whilst simultaneously affecting the outcome of the treatment. Any such predictors act as a confounders between treatment decision (as one of the predictor variables) and treatment outcome.
In such cases, we must be extra cautious as we risk
unwanted bias in the causal effect estimator \citep{rosenbaum83},
if we ignore such correlations.
Several authors have tackled the presence of
confounder variables. 
\Citet{Robins1986ANA} used a graphical
approach to identify the causal parameters.
\Citet{rosenbaum1985} suggested a link model to estimate
the propensity scores for all individuals. Subsequently, several other
methods have been proposed based on propensity score matching;
see \citep{winship99,stuart10} for a brief review.

One of the earlier Bayesian approaches to causal inference
can be found in \citep{rubin1978}. More recently,
with the rise of high dimensional data,
Bayesian methodologies have grown in popularity.
\Citet{Crainiceanu2008} proposed a bi-level 
Bayesian model averaging based method for estimating the causal 
effect. \Citet{wang2015} suggested BAC (or, Bayesian adjustment for
confounding),
where an informative prior obtained from
the treatment model is applied on the outcome model for
estimating the causal effect. Several other methods were
proposed to tackle confounders,
see for instance \citep{Zigler2014,Hahn2018} among others \added{for
	a Bayesian perspective and
	\citep{zhang_2018} for a survey of methods for addressing unmeasured confounding.}

In this paper, we take inspiration from the approach of \citet{koch2020}, who proposed a bi-level spike and slab prior for causal effect estimation \added{in high dimensional problems (i.e.~when number of predictors
	is larger than the number of observations)}. 
	They considered a data-driven adaptive approach to
propose their prior which reduces the variance of the causal estimate. 
Our approach \added{however focuses on the other aspect of high
dimensional causal inference problem, ie.~variable selection. To achieve that we rely} on \added{prior}
sensitivity analysis, where instead of using a single prior, 
we consider a set of priors \citep{BERGER1990303}. 
\added{Prior sensitivity analysis for causal inference has
been a topic of interest lately.} \citet{zaffalon20a} 
used credal networks in structured causal models
for causal inference; \citet{raices_cruz22} \added{performed 
	a meta analysis in a robust Bayesian framework for causal effect estimation. However, variable selection in causal estimation
	problem has not been investigated in robust Bayesian framework.} 
This motivates us to \added{investigate the role and applicability of} prior sensitivity analysis in high dimensional \added{causal estimation} problems. This is particularly beneficial, as
in high dimensional problems,
we have to rely on very limited \added{observations} to perform our Bayesian analysis and
\added{as a result variable selection} with a single prior
can be unreliable \citep{basu2023robust} \added{in many cases, Moreover, in causal effect estimation, failing to correctly identify a relation between the treatment effect and predictor can lead to
harmful side-effects. Therefore, it
is extremely important to adopt a cautious approach in selecting
or rejecting a variable.}
To \added{achieve this cautious paradigm and to} perform a prior sensitivity analysis, we rely on expert opinion to elicit a set of priors 
based on empirical evidence. 
This allows us to construct the problem of 
predictor selection
in a framework where abstention has a relatively positive gain i.e.~when the cost of further tests/data collection is
\added{lower than that of incorrectly treating} a subject. 

Our framework considers a set of continuous spike and slab priors 
\citep{ishwaran2005} for
predictor selection.
We thereby construct a Bayesian group LASSO \added{(least absolute shrinkage and selection operator)} \citep{xu2015} type problem.
To perform sensitivity analysis,
we consider a set of beta priors on the covariate selection 
probability of the spike and slab priors. We use the posteriors of this
covariate selection probability for identifying the active predictors. Finally, 
we consider a post-hoc coefficient adjustment method \citep{hahn2015}
to recover sparse estimates associated with either the outcome or the
treatment model. 

The rest of the paper is organised as follows. In \cref{sec:causal}
we give a formal description of the causal estimation problem in the
context of linear regression. \Cref{sec:bayes} is focused on the
Bayesian analysis of causal inference problems, followed by the
motivation of a robust Bayesian analysis along with our proposed decision 
theoretic framework for predictor selection. In \cref{sec:sim}, 
we provide results of simulation studies under different scenarios 
and show the possible applications in real life problems. Finally, 
we discuss our findings and conclude this paper in \cref{sec:conc}.

\section{Causal Estimation}\label{sec:causal}

Let an observational study on $n$ individuals give us
\emph{treatment outcomes} $Y\coloneqq(Y_1, \dots, Y_n)$ with 
corresponding \emph{treatment decisions} $T\coloneqq(T_1, \dots, T_n)$.
Here, we use an indicator to represent the treatment decision. That is, $T_i$ is $1$ if the $i$th patient was treated, and $0$ otherwise. Similarly,
$Y_i$ is the treatment outcome of the $i$th patient, represented as some real-valued quantity.

Regression methods are widely used in causal effect estimation. The
main idea behind these regression methods is to remove the
correlation between the treatment indicator and the error term
\citep{winship99,HECKMAN1985}.
To do so, we rely on $p$ observed quantities, called
\emph{predictors}, denoted by $X\coloneqq$ $[X_1^{\top}, \dots, X_n^{\top}]^{\top}$
where each $X_i\in\mathbb{R}^p$.
Each $X_i$ is treated as a $p$-dimensional row vector,
so $X$ is a $n\times p$ matrix.
Now, let
$\beta \coloneqq (\beta_1$, \dots, $\beta_p)^{\top}$ denote the vector of regression
coefficients
related to the predictors, and let $\beta_T$ denote a regression coefficient related to the
treatment decision.
\added{%
Following the usual approach in the literature
(see for instance \citep{winship99,HECKMAN1985}),
we model the outcome using a linear model%
}
\begin{equation}
	Y_i =  T_i \beta_{T} + X_i\beta \added{+ \beta_0} + \epsilon_i
\end{equation}
where $\epsilon_i\sim \mathcal{N}(0, \sigma^2)$,
independent of $T_i$ and $X_i$.
Note that both $T_i$ and $X_i$ are predictors for $Y_i$ in the above model.
However, when we talk about predictors in this paper, we usually mean just the components of $X_i$.

To decide whether or not to treat a new individual with given predictors,
we are mainly interested in the effect of the treatment on the outcome.
More precisely, the causal effect of a new individual, indexed as $n+1$,
whose outcome $Y_{n+1}$ is not yet observed, and with observed predictors $X_{n+1}=x_{n+1}$, is defined by:
\begin{align}
  \begin{split}
  \delta(x_{n+1})
    &\coloneqq\mathbb{E}(Y_{n+1}\mid X_{n+1}=x_{n+1},T_{n+1} =1) \\
    &\qquad - \mathbb{E}(Y_{n+1}\mid X_{n+1}=x_{n+1},T_{n+1}=0)
  \end{split}
  \intertext{For our model, due to linearity of expectation, we have that}
  \begin{split}
  \delta(x_{n+1})
    &=\beta_T
      +x_{n+1}\beta+\mathbb{E}(\epsilon_{n+1}\mid X_{n+1}=x_{n+1},T_{n+1} =1) \\
    &\qquad\,\,\, %%% for alignment
      -x_{n+1}\beta-\mathbb{E}(\epsilon_{n+1}\mid X_{n+1}=x_{n+1},T_{n+1} =0)
  \end{split}
\intertext{and because $\epsilon_{n+1}$ is independent from $X_{n+1}$ and $T_{n+1}$,}
  &=\beta_{T}.
\end{align}
Note that, for this model, the causal effect $\delta(x_{n+1})$
does not depend on the observed value $x_{n+1}$ of $X_{n+1}$.
So, to find the causal effect, we simply need to estimate $\beta_T$.
\added{%
Note that, if interaction terms between $X_i$ and $T_i$ were present in
the model (for example a term of the form, say, $T_iX_i\eta$ for some
parameter vector $\eta$), that would result in a dependence of the
causal effect on $x_{n+1}$.%
}

To estimate $\beta_T$ from the data $X$, $Y$ and $T$,
especially in the presence of confounders,
we also need to consider the
association between the treatment indicators $T$ and the predictors $X$.
A common choice in the literature is to use a probit link function
\citep{winship99},
\added{%
though other link functions, such as the logit, can also be used
\citep{HECKMAN1985}.%
}
In this way, we can
specify the conditional probability that subject $i$ receives the treatment through a linear model. 
That is, for another vector of regression coefficients 
$\gamma\coloneqq(\gamma_1, \cdots, \gamma_p)^{\top}$ we
assume
\begin{align}
	P(T_i=1\mid X_i) = \Phi(X_i\gamma\added{+\gamma_0})
\end{align}
where $\Phi$ denotes the cumulative distribution function
of a standard normal distribution.
\added{%
The key assumption made here is that
there is a monotone relationship between
the predictors and the probability of treatment.
Here too, interaction terms between the $X_i$ could be added
to form more complex models if so desired.%
}

To incorporate this probit
link function, we model the $T_i$ as follows \citep{albert93}:
\begin{align}
    T_i^* &= X_i\gamma\added{+\gamma_0} +u_i \\
    T_i   &= \mathbb{I}(T_i^*>0)
    =
    \begin{cases}
    1 & \text{if }T_i^*>0 \\
    0 & \text{otherwise}
    \end{cases}
\end{align}
where $u_i\sim\mathcal{N}(0,1)$.
With this model, indeed
\begin{align}
  P(T_i=1\mid X_i)
  &=P(T_i^*>0)=P(u_i>-X_i\gamma\added{-\gamma_0})=1-P(u_i\le -X_i\gamma\added{-\gamma_0}) \\
  &=1-\Phi(-X_i\gamma\added{-\gamma_0})=\Phi(X_i\gamma\added{+\gamma_0}).
\end{align}

Now, to construct the joint likelihood function, we define an extended
output $2n\times 1$ column vector
$W\coloneqq\left(\begin{smallmatrix}Y \\ T^*\end{smallmatrix}\right)$
and corresponding $2n\times(2p+\added{3})$ dimensional design matrix
\begin{align}
	Z &\coloneqq
        \begin{bmatrix}
           T_1 & X_1 & \added{1} & 0 & \added{0} \\
           \vdots & \vdots & \added{\vdots} & 0 & \added{0} \\
           T_n & X_n & \added{1} & 0 & \added{0} \\
           0 & 0 & \added{0} & X_1 & \added{1} \\
           \vdots & \vdots & \added{\vdots} & \vdots & \added{\vdots} \\
           0 & 0 & \added{0} & X_n & \added{1}
        \end{bmatrix}
        =
	\begin{bmatrix}
		X_O & 0 \\
		0 & X_T
	\end{bmatrix}
\end{align}
where, $X_O \coloneqq [T, X, \added{\mathbf{1}_n}]$ and $X_T \coloneqq [X, \added{\mathbf{1}_n}]$. Then, considering the assumption of
Gaussian error terms, we have the following likelihood distribution
\begin{align}
	W\mid Z, \beta_T, \beta, \added{\beta_0,} \gamma, \added{\gamma_0,} \sigma^2 \sim\normal\left(Z\nu, \Sigma\right)\label{eq:like:group},
\end{align}
where $\nu \coloneqq (\beta_T, \beta^{\top}, \added{\beta_0,} \gamma^{\top}\added{, \gamma_0})^{\top}$ and
\begin{align}
	\Sigma &\coloneqq
	\begin{bmatrix}
		\sigma^2{I}_n & 0 \\
		0 & {I}_n
	\end{bmatrix}.
\end{align}

\section{\added{Robust} Bayesian Causal Estimation}\label{sec:bayes}

The likelihood given by \cref{eq:like:group} gives us
a foundation for a Bayesian group LASSO 
\citep{xu2015} type model. In this way, we can look into the posterior selection
probability of each predictor. \added{In this section, we formally introduce our
	proposed methodology for causal estimation and we call it as
	`robust Bayesian causal estimation' as we perform a robust
	Bayesian analysis \citep{BERGER1990303} to achieve a cautious
	variable selection paradigm. There are several
ways to construct spike and slab priors for
variable selection. In our method}, we consider a continuous type
prior \citep{basu2023robust,ishwaran2005} for faster posterior
computation.

\subsection{Hierarchical model}

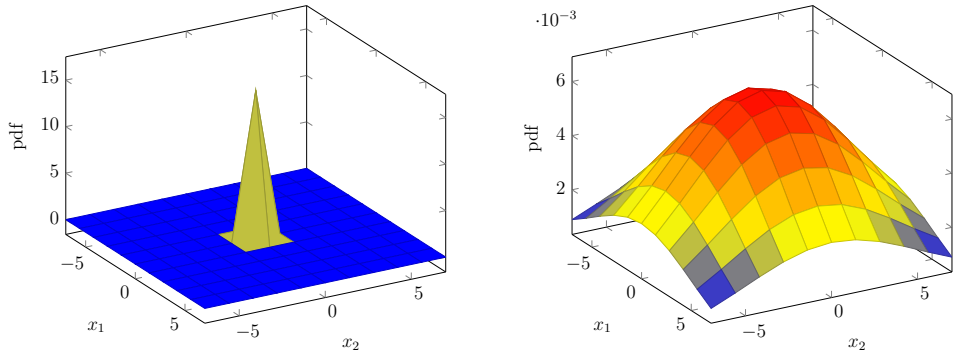
\begin{figure}[h]
	\begin{center}
\begin{tikzpicture}[scale=0.6]
    \begin{axis}[
        width=10cm,
        view={60}{30},
        xlabel={$x_1$},
        ylabel={$x_2$},
        zlabel={pdf},
        ]
        \addplot3[
            surf,
            samples=11,
            domain=-7:7,
            y domain=-7:7,
        ] 
        {exp(-0.5*(x^2 + y^2)/0.01) / (2*pi*0.01)};

    \end{axis}
\end{tikzpicture}
\qquad
\begin{tikzpicture}[scale=0.6]
    \begin{axis}[
        width=10cm,
        view={60}{30},
        xlabel={$x_1$},
        ylabel={$x_2$},
        zlabel={pdf},
        ]
        \addplot3[
            surf,
            samples=11,
            domain=-7:7,
            y domain=-7:7,
        ] 
        {exp(-0.5*(x^2 + y^2)/25) / (2*pi*25)};

    \end{axis}
\end{tikzpicture}
	\end{center}
	\caption{\added{Spike (left) and slab (right) components of a bivariate distribution for $\tau_0 = 0.1$, $\tau_1 = 5$ and $\sigma=1$.}}
	\label{fig:ssbl}
\end{figure}

Let $\pi_j$ denote the prior probability that the $j$-th
predictor is associated with the outcome or the 
treatment. That is, conceptually,
\begin{equation}
	\pi_j \coloneqq P\left((\beta_j,\gamma_j)\not=(0,0)\right).
\end{equation}
Practically, we model this by defining the following hierarchical model
so that,
for $1\le j\le p$,
\begin{align}
	\label{eq:spike:slab:prior:beta:gamma}(\beta_j,\gamma_j)^{\top} \mid \pi_{j}, \sigma^2 &\sim 
	\pi_{j}\normal\left( \begin{bmatrix}
		0 \\
		0
	\end{bmatrix}, 
	\tau_1^2\begin{bmatrix}
		\sigma^2 & 0 \\
		0 & 1
	\end{bmatrix}\right)
	+ (1-\pi_{j}) \normal\left(\begin{bmatrix}
		0 \\
		0
	\end{bmatrix}, 
	\tau_0^2\begin{bmatrix}
		\sigma^2 & 0 \\
		0 & 1
	\end{bmatrix}\right)\\
	\beta_T\mid \sigma^2 &\sim \normal\left(0, \sigma^2\right)\label{eq:prior:causal}\\
        \beta_0\mid \sigma^2 &\sim \normal\left(0, \sigma^2\right)\label{eq:prior:int:out}\\
        \gamma_0 &\sim \normal\left(0, 1\right)\label{eq:prior:int:trt}\\
        \frac{1}{\sigma^2}&\sim \text{Gamma}(a, b)\\
	\pi_{j} &\sim\text{Beta}\left(sq_j, s(1-q_j)\right).
\end{align}
In the hierarchical model, we fix sufficiently small $\tau_0$
$(1\gg\tau_0>0)$ so that  $(\beta_j, \gamma_j)$ has its probability mass 
concentrated around zero. Therefore, this represents the spike component of our prior specification. 
For the slab component, we consider $\tau_1$ to be large so that $\tau_1\ge 1$. This allows the prior for $(\beta_j,\gamma_j)$ to be flat beyond the spike component at the origin. 
We illustrate the components of a bivariate spike and slab prior in 
\cref{fig:ssbl} (with fixed $\sigma=1$). We generate the spike component 
with $\tau_0=0.1$ and the slab component with $\tau_1=5$.

For the precision term $1/\sigma^2$, a natural choice of prior is the gamma distribution
as it allows the control of both the location and the scale of the precision.
To ensure that the prior is able to represent the data, we consider $b=1$ and 
fix $a$ so that it represents the prior mean of the precision.
Alternatively, when $b=1$, we know that the interval
$[0, 3a]$ contains the true value of the precision parameter with probability close to $0.95$.
So, we can also use a prior judgement on the 95\% quantile to set $a$.
We use a beta prior to
model our uncertainty about
the selection probabilities $\pi_j$ where  $q_j$ represents our prior expectation of $\pi_j$ and $s$ acts as 
a concentration parameter.
For the causal effect \added{in \cref{eq:prior:causal} and intercept term of the outcome model in \cref{eq:prior:int:out}}, we want to use a Gaussian distribution that 
matches the scale of the noise term. Therefore, we consider $\beta_T\added{, \beta_0\mid\sigma^2}\sim \normal(0,\sigma^2)$. \added{Similarly, for the intercept of the treatment model we match the scale of the probit model and consider $\gamma_0\sim \mathcal{N}(0,1)$}.

\begin{figure}[h]
	\centering
	\begin{tikzpicture}[params/.style={circle, draw=black!60, very thick, minimum size=7mm},
		hyper/.style={circle, draw=black!60, fill=black!20, thick, minimum size=7mm},
		post/.style={circle, draw=black!60, fill=green!20, thick, minimum size=7mm},
		latent/.style={rectangle, draw=black!60, fill=black!10, dashed, minimum size=7mm},
		data/.style={rectangle, draw=black!60, thick, minimum size=8mm}]
		\node[params] (1) at (0,0) {$\pi$};
		\node[data] (2) at (3,0) {$X$};
		\node[data] (3) at (6,0) {$T$};
		\node[params] (4) at (1.5,1.5) {$\gamma$};
		\node[params] (15) at (1.5,0) {$\gamma_0$};
		\node[latent] (5) at (4.5,1.5) {$T^*$};
		\node[params] (6) at (1.5,-1.5) {$\beta$};
		\node[params] (16) at (1.5,-3) {$\beta_0$};
		\node[data] (7) at (3.5,-1.5) {$Y$};
		\node[params] (8) at (0,-3) {$\sigma^2$};
		\node[params] (10) at (6,-1.5) {$\beta_T$};
		\node[hyper] (11) at (-1.5,-.8) {$s$};
		\node[hyper] (12) at (-1.5,.8) {$q$};
		\node[hyper] (13) at (-1.5,-2.2) {$a$};
		\node[hyper] (14) at (-1.5,-3.8) {$b$};
		\draw[black, dashed] (0.75,-3.7) rectangle (7,2.1);
		
		\path (1) edge[->]  (6);
		\path (1) edge[->]  (4);
		\path (8) edge[->]  (6);
		\path (6) edge[->]  (7);
		\path (2) edge[->]  (7);
		\path (2) edge[->]  (5);
		\path (5) edge[<-] (4);
		\path (5) edge[->] (3);
		\path (3) edge[->]  (7);
		\path (10) edge[->]  (7);
		\path (2) edge[dashed][->] (3);
		\path (8) edge[bend right = 60][->]  (10);
		\path (11) edge[->]  (1);
		\path (12) edge[->]  (1);
		\path (13) edge[->]  (8);
		\path (14) edge[->]  (8);
		\path (15) edge[->]  (5);
		\path (16) edge[->]  (7);
		
	\end{tikzpicture}
	\caption{Probabilistic graphical representation for causal inference with \added{our} Bayesian hierarchical model.}
	\label{fig:regress}
\end{figure}
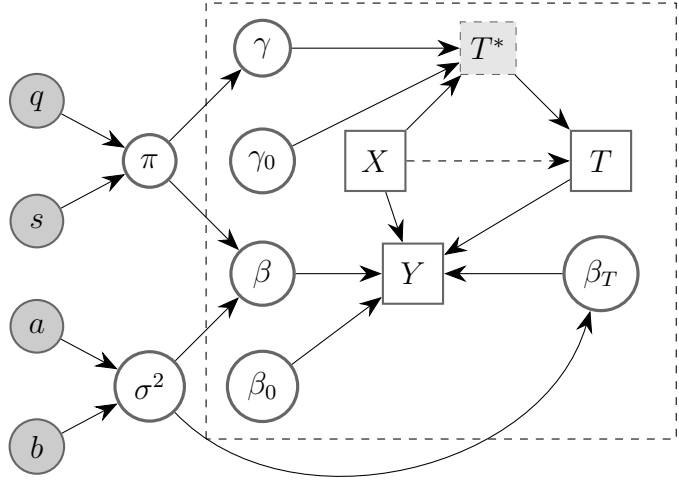

In \cref{fig:regress}, we show a probabilistic graphical representation
of our hierarchical model. In the figure, grey circular nodes represent the
prior hyper-parameters which will be used for sensitivity analysis
of the model. The transparent circular nodes are used to denote
the modelling parameters which are our quantities of interest. 
The observed quantities are denoted with transparent rectangular
nodes. We also use a grey rectangular node to denote the intermediate
latent variable $T^*$. We use directed edges to denote the
relationship between different nodes. However, we use a dashed
edge between $X$ and $T$ as they are related through the latent
variable $T^*$. 

\subsection{Robust Bayesian Analysis}
\label{sec:robustbayesanalysis}

The hierarchical model presented above is a standard spike and slab model for
variable selection and performs well when we have sufficient data. 
However, especially in many situations, we do not always have sufficient data. Moreover, 
we also must be cautious about the side effects of a treatment.
Therefore, we are particularly interested in constructing a robust Bayesian framework
for variable selection. In this way, when we are preparing guidelines for treatment, we 
can have the option to ask for more data before reaching any conclusion. To achieve this,
we consider a utility based framework with three possible \added{outcomes}.

In particular, through predictor selection, we also want 
to check if a certain bio-marker should be considered
for the treatment decision. For example, say we can observe blood pressure with some
other bio-markers and want to decide whether our
treatment guideline should also consider the blood
pressure of the subject before treating them. This is
useful as an unnecessary treatment of a subject can have severe consequences because of the medicinal
side effects. In general, it is hard to associate such consequences 
with a suitable loss function. Instead, we
assume that we can always revert any initial incorrect treatment by further treatments, 
and we can associate a loss function with the cost of further treatments.
So, we will associate two constant loss values $\ell_1$ and $\ell_2$ 
with false positives (falsely selected predictor) and 
false negatives (falsely rejected predictor) respectively. 
Clearly, false positives may lead to unwanted side effects and
false negatives may lead to incorrect treatment of the patient. Finally, we associate
a loss value $\ell_3$ for abstention from selecting
a variable which can be interpreted as the cost of further tests to determine whether that bio-marker
is important for constructing the treatment guideline.
Ideally, in most cases, $\ell_3\ll \ell_1,\ell_2$. However, in certain scenarios,
this might not be the case, especially when the condition of a subject deteriorates rapidly
over time.

Now, based on this notion of abstaining from selecting a predictor, we can perform
a sensitivity analysis over a set of priors on the prior selection probability.
That is, we can consider a set of possible values for $q$ such that
$q\in\mathcal{P}$, where $\mathcal{P} \subseteq \left(0, 1\right)^{p}$.
Here, the equality occurs for the near vacuous case. However, in real-life
situations, performing a robust Bayesian analysis for the near vacuous case is 
not practical. Instead, we incorporate expert elicitation to define our model.

\added{%
For instance, assume $\underline{k}$ and $\overline{k}$ represent the expert's bounds of the prior expectation on the
total number of variables present in either of the models. We can then consider $\mathcal{P}=\left[\underline{k}/p, \overline{k}/p\right]^p$.
Using an interval for the prior expectation on the total number of active variables
gives us a more cautious approach to specifying the prior distribution on variable selection,
and thus more robust inferences.%
}

\added{%
Alternatively, we may also use the empirically observed correlations from the data directly. This is particularly common in ultra high
dimensional problems (when $p\gg n$) for reducing the dimensionality
of the problem \citep{fan2008}. We can also use this approach
to have a better prior judgement since any predictors that are correlated with the outcome are good candidates to be active.
When doing so, we need
% \added{
a prior judgement on what is a reasonable
%}
correlation between active predictors and the outcome.
Say the expert judges that
an active predictor has a correlation with the outcome
that lies typically in $[-1,-c]\cup[c,1]$, i.e.
an absolute correlation larger than $c$.
Let $k_c$ be the number of predictors with absolute marginal correlation greater than $c$.
We could then consider $q=k_c/p$ for the prior, as it gives a prior estimate on the selection probability that is consistent with a prior predictive expectation of $k_c$ active variables.
Now, it is in general quite difficult to specify an exact value for $c$ a priori.
Therefore, we consider an interval $[\underline{c},\overline{c}]$ for $c$, leading to
$\mathcal{P}=\left[k_{\overline{c}}/p, k_{\underline{c}}/p\right]^p$
(note that $k_c$ is monotonically non-increasing in $c$).
}

\paragraph{Variable selection}
\added{Ideally, we should check the joint posterior probability of $\pi_j$'s to select the most probable model. However, this means we have to search a space of dimension $2^p$, which is practically impossible when $p$ is very large. Instead, we can use the posterior of individual $\pi_j$ as \citet{barbieri2004} showed that median probability model gives the optimal model. That is we can set a threshold of $1/2$ to select a variable. Therefore,}
we consider the $j$-th predictor to be removed from both the
treatment and outcome model, if
\begin{align}\label{eq:vs:remove}
	\uexp (\pi_j\mid W)\coloneqq \sup_{q\in \mathcal{P}} \mathbb{E}_q(\pi_j\mid W) < 1/2.
\end{align}
Similarly, we consider the $j$-th predictor to be present in at least one of the models, if
\begin{align}\label{eq:vs:sel}
	\lexp (\pi_j\mid W)\coloneqq \inf_{q\in \mathcal{P}} \mathbb{E}_q(\pi_j\mid W) \ge 1/2.
\end{align}
Otherwise, we consider the variable to be indeterminate,  in which case we abstain from putting
it in any of the models but instead just report a lack of information.

\subsection{\added{Coefficient Adjustment and Refit}}
In general, \added{our} framework is
intended
for \added{robust} variable selection \added{in causal effect estimation problem}. However, \added{one might also be interested in}
model fitting and prediction, \added{for that} we need to evaluate the values 
of the regression coefficients. To do so, we first need to find the set of active
predictors with respect to our prior expectation of the selection probability $q$.
For any fixed $q$, we define the set $S(q)$ as the set of all variables which are active
in the treatment model or in the outcome model:
\begin{equation}
	S(q)\coloneqq
	\left\{j\colon \mathbb{E}_q(\pi_j\mid W) \ge 1/2\right\}.
\end{equation}
For sensitivity analysis,
the intersection of $S(q)$ over all $q$ gives us the set of
active variables obtained through \cref{eq:vs:sel}.
Similarly, the union gives us the set of
variables that are not removed through \cref{eq:vs:remove}.
That is:
\begin{align}
    \mathcal{S}_*&\coloneqq \left\{j:\lexp (\pi_j\mid W)\ge1/2\right\}
    = \bigcap_{q\in \mathcal{P}}S(q), \\
    \mathcal{S}^*&\coloneqq \left\{j:\uexp (\pi_j\mid W)\ge1/2\right\}
    = \bigcup_{q\in \mathcal{P}}S(q).
\end{align}
Clearly, $\mathcal{S}_*\subseteq\mathcal{S}^*$.
$\mathcal{S}_*$ represents the set of variables that are sure to be selected,
$\{1,\dots,p\}\setminus\mathcal{S}^*$ represents the set of variables that are sure to be removed, and
$\mathcal{S}^*\setminus\mathcal{S}_*$ represents the set of variables about which we are undecided.
In this way, through sensitivity analysis, our approach incorporates robustness.

We can derive bounds on the posterior means of the parameters as follows:
\begin{align}
\label{eq:beta:lower}
\underline{\beta}_j&\coloneqq\lexp (\beta_j\mid W)= \inf_{q\in \mathcal{P}} \mathbb{E}_q(\beta_j\mid W) \\
\label{eq:beta:upper}
\overline{\beta}_j&\coloneqq\uexp (\beta_j\mid W)=\sup_{q\in \mathcal{P}} \mathbb{E}_q(\beta_j\mid W)
\end{align}
with similar expressions for 
$\underline{\beta}_T$, $\overline{\beta}_T$,
$\underline{\gamma}_j$ and $\overline{\gamma}_j$.
If we take the posterior expectation interval $[0,0]=\{0\}$ on a regression coefficient to represent absence of a variable, then our bounds on the regression coefficients are generally not sparse, because we use continuous 
spike and slab priors.

Moreover, with our variable selection we only determine whether the variable 
is included in at least one of the models.
To determine which predictors
influence the outcome ($\beta_j\neq 0$),
the treatment ($\gamma_j\neq 0$), or both,
and to understand the degree of assocation
(i.e.\ the magnitude of $\beta_j$ and/or $\gamma_j$),
we apply the 
``decoupled shrinkage and selection'' (DSS) method proposed by \citep{hahn2015}. 
For that, we solve the following adaptive LASSO-type \citep{Zou2006}
problems:
\begin{align}
	\hat{\beta}^D_{S(q)} &= 
	\arg\min_{\beta_{S(q)}} \frac{1}{n}\|X_{S(q)}\hat{\beta}_{S(q)}
	- X_{S(q)} \beta_{S(q)}\|_2^2 + \lambda\sum_{j\in S(q)} 
	\frac{|\beta_{j,S(q)}|}{|\hat{\beta}_{j,S(q)}|}
\end{align}
and
\begin{align}
	\hat{\gamma}^D_{S(q)} &= 
	\arg\min_{\gamma_{S(q)}} \frac{1}{n}\|X_{S(q)}\hat{\gamma}_{S(q)}
	- X_{S(q)} \gamma_{S(q)}\|_2^2 + \lambda\sum_{j\in S(q)} 
	\frac{|\gamma_{j,S(q)}|}{|\hat{\gamma}_{j,S(q)}|}
\end{align}
where $q\in \mathcal{P}$,
where $\hat{\beta}_{S(q)}$ 
and $\hat{\gamma}_{S(q)}$ are the posterior means 
of the regression coefficients with respect to
the predictors that belong to $S(q)$.
By varying $q$, this gives us a set of point estimates for the model parameters $\beta$ and $\gamma$, along with a more detailed selection of individual $\beta_j$ and $\gamma_j$.

To compute the posterior bounds (as in \cref{eq:vs:remove,eq:vs:sel,eq:beta:lower,eq:beta:upper}), unfortunately, we usually have to resort to brute force optimisation, due to the lack of tractable expressions for the posterior expectations. 
This is obviously a major drawback of this approach.

\paragraph{Refit}

In our setting, the DSS method only gives us a set of point estimates for the final selection of variables: some coefficients may be always selected, some never, and some will be indeterminate. For the final inference model, the modeller will need to make a judgement about which of the indeterminate coefficients $\beta_j$ and $\gamma_j$ to include in the final model or not. Once done so, the model can be refitted to account for the effect of variable selection on the estimation of the model parameters.

To do so, we can again use our Bayesian model without $\pi_j$ (as there is no selection anymore), and with priors
\begin{align}
\beta_j\mid\sigma^2&\sim\mathcal{N}(0,\tau_1^2\sigma^2) \\
\gamma_j\mid\sigma^2&\sim\mathcal{N}(0,\tau_1^2)
\end{align}
for those $\beta_j$ and $\gamma_j$ that are selected in the model, with the remaining $\beta_j$ and $\gamma_j$ set to zero.
This is similar to the spike and slab prior from \cref{eq:spike:slab:prior:beta:gamma} but without the spike component.

We expect this to have only a small effect on the mean and variance of the estimated parameters. This refit is useful to validate the variable selection and to improve the estimating of the model parameters, including the causal effect $\beta_T$. Indeed, since there are fewer parameters for the same data, the estimates are expected to have less uncertainty.

Note that here, we described a precise Bayesian refit model, but obviously this could be extended to robust Bayesian refit models too.

\section{Simulation Studies}\label{sec:sim}

For the simulation studies, we consider \added{2 different cases each with 2 sub-cases, amounting to 4 studies in total}. In each
\added{of these 4 studies}, we generate the design matrix $X$ such that $X_i\sim\mathcal{N}(0, \Sigma)$
for $1\le i\le n$ where $\Sigma_{ij} = 0.3^{|i-j|}$. In this way, we 
generate predictors for our model with mild correlations between them.
We then use the following distributions to generate the outcome and
the treatment indicator: 
\begin{equation}
    T_i \sim \text{Bernoulli}\left(1/(1+\exp(-X_i\gamma))\right)
    \quad\text{and}\quad
    Y_i = 4T_i + X_i\beta + \epsilon_i.
\end{equation}
where $\epsilon_i\sim\mathcal{N}(0,0.1^2)$.
Note that the simulated causal effect $\beta_T$ is equal to $4$.

\added{In case 1, we consider an} increasing number of observations. \added{We} have two
sub-cases: in \added{case 1a} we consider all active variables to be confounders
and in \added{case 1b} we consider some active variables which are only related to
the outcome model. 
\begin{description}
    \item[\added{Case} 1a] --- $|\gamma_j|, |\beta_j|>0$ for $j\le 10$
    \item[\added{Case} 1b] --- $|\gamma_j|>0$ for $j\le 10$ and $|\beta_j|>0$ for $j\le 15$
\end{description}
For both \added{case 1a and 1b}, we consider different numbers of observations $n$ where
\added{$n=20+ 5k$ for $k=1,2,\dots,11$ and $p=50$}
predictors. This way, we check the efficiency of our method with varying level of information.

For \added{case 2}, we check our method for varying number of predictors 
(and hence sparsity level, i.e.\ the percentage of active variables present in the model).
Similar to \added{case 1} we also have two \added{sub-cases}:
\begin{description}
    \item[\added{Case} 2a] --- $|\gamma_j|, |\beta_j|>0$ for $j\le 10$
    \item[\added{Case} 2b] --- $|\gamma_j|>0$ for $j\le 10$ and $|\beta_j|>0$ for $j\le 15$
\end{description}
For both
\added{case 2a and 2b}, we consider different numbers of predictors $p$ where
\added{$p=20+ 5k$ for $k=1,2,\dots,11$ and $n=40$}
subjects.

\added{For all four cases, we consider 20 replicates for an empirical statistical analysis to check the consistency and robustness of our approach.}

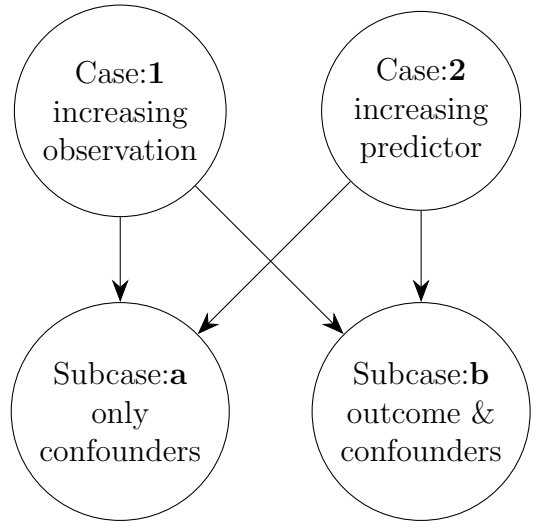
\begin{figure}
	\begin{center}
		\begin{tikzpicture}[node distance=2cm, auto]
			\node (1) at (0,2) [circle,draw,align=center] {Case:\textbf{1}\\increasing \\ observation};
			\node (2) at (4,2) [circle,draw,align=center] {Case:\textbf{2}\\increasing \\ predictor};
			\node (a) at (0,-2) [circle,draw,align=center] {Subcase:\textbf{a}\\only \\ confounders};
			\node (b) at (4,-2) [circle,draw,align=center] {Subcase:\textbf{b}\\outcome \& \\ confounders};
			\draw[->] (1) to (a);
			\draw[->] (1) to (b);
			\draw[->] (2) to (a);
			\draw[->] (2) to (b);
			%\draw[->,dashed] (predictor) to (outcome);
		\end{tikzpicture}
	\end{center}
	\caption{\added{Diagram of the simulation setup for comparative study.}}
	\label{fig:experiment}
\end{figure}

\added{We use these studies to compare our method with three other approaches. From now on, for the sake of illustration,}
we use the following acronyms: RBCE for 
robust Bayesian causal estimation (our method); SSCE for spike and
slab causal estimation \citep{koch2020}; BSSCE for bi-level spike and slab causal
estimation \citep{koch2020}; and BSSL for Bayesian spike and slab LASSO
\citep{xu2015}.

\paragraph{Metrics}

\added{ 
	As mentioned earlier, to perform our statistical analyses we use 20
	replications. To evaluate the accuracy of estimation, we consider
	mean and median values obtained from these 20 samples. Similarly, to check the
	dispersion, we use standard deviation (denoted by sd), mean squared 
	error with respect to the true value (denoted by MSE) and coverage (percentage) of the true value within the
	95\% posterior credible interval (denoted by CI\%). Finally, 
	to check the accuracy of variable selection, we evaluate false average positive numbers (denoted by FP), average false negative numbers (denoted
	by FN) and average number of indeterminate variables (denoted by ID). Clearly
	for other classical methods ID is equal to zero and therefore 
	is not presented in the tables. We also define a misspecification loss
	in the following way for illustration :
	\begin{equation}
		\text{Misspecification Loss} = \frac{FP}{TN} + \frac{FN}{TP} + 0.2* \frac{ID}{\text{Total no of predictors}}.
	\end{equation}
}

\added{Note that since RBCE gives interval estimates, CI\% is calculated using the minimum of the lower bounds of credible intervals and the maximum of the upper bounds of credible intervals.}

\paragraph{Elicitation}

\added{%
To elicit $\mathcal{P}$,
as discussed earlier in \cref{sec:robustbayesanalysis},
we use the empirically observed correlations from the data directly.
For expert elicitation, we follow the correlation guidelines mentioned in \citep{AKOGLU201891}
where the authors provide \cref{tab:corr}.}

\begin{table}
	\centering
	\caption{\added{Interpretation of the Pearson's and Spearman's correlation coefficients in absolute values}}
	\begin{tabular}{|c|l|l|l|}
		\hline
		Absolute & Dancey \& Reidy\citep{dancey2007statistics} & Chan YH\citep{chan2003biostatistics} & Quinnipiac University \\
		Correlation & (Psychology)& (Medicine) & (Politics)\\
		\hline
		1.0 & Perfect & Perfect & Perfect \\
		0.9 & Strong & Very strong & Very strong \\
		0.8 & Strong & Very Strong & Very strong \\
		0.7 & Strong & Moderate & Very strong \\
		0.6 & Moderate & Moderate & Strong \\
		0.5 & Moderate & Fair & Strong \\
		0.4 & Moderate & Fair & Strong \\
		\cdashline{2-2}
		0.3 & Weak & Fair & Moderate \\
		0.2 & Weak & Poor & Weak \\
		\cdashline{4-4}
		0.1 & Weak & Poor & Negligible\\
		0.0 & Zero & Zero & Zero \\
		\hline
	\end{tabular}
	\label{tab:corr}
\end{table}

\added{From \cref{tab:corr}, we notice that the number of
	labelled relations is different for different columns. As
	a result it is difficult to obtain a single value for $c$
	as mentioned in \cref{sec:robustbayesanalysis}. Instead,
	we can disregard the last labelled relation (other than zero)
	and assume that $c$ is typically larger than a value lying in the interval $[0.15,0.35]$
	Let $\overline{k}$ be the number of predictors with absolute marginal correlation greater than $0.15$
and let $\underline{k}$ be number of predictors with absolute marginal correlation greater than $0.35$.
Then $\mathcal{P}=[\underline{k}/p , \overline{k}/p]^p$ gives us a prior bound on the selection probability of each predictor, reflecting our prior expert judgement.%
}

\paragraph{Initialisation} 
To implement our method, we use \texttt{rjags} \added{\citep{rjags2023}} and for the other three
methods we use the code provided in the appendix of \citep{koch2020}.
\added{However, we modify to accommodate analysis with `high dimensional'
	data.}
For our method, we set $\tau_0=10^{-6}$ and $\tau_1=1$ to construct the
spike and slab prior.
%%% MT to TB: where does a=10 come from? can we say something about that?
For the noise term, we set $a=50$ and $b=1$.
To perform 
our Bayesian analysis with \texttt{rjags}, \added{we discard 500 burn in samples and consider 2500 MCMC samples} to compute the
posterior estimates. For the other methods we use the in-built settings 
to initiate the analyses. \added{We also transform the data so that the data is centred around $0$
for the outcome model to avoid having an intercept term.}

\begin{table}[ht]
	\tiny
	\centering
	\caption{\added{Comparison of different methods for varying number of observations where all the active variables are confounders.}}
	\label{tab:causal1a}
	\subcaption{Accuracy in estimation of causal effect}
	\begin{tabular}{l|cccc|cc|cc|cc}
		\hline
		\multicolumn{1}{l|}{}&
		\multicolumn{4}{c|}{RBCE}&
		\multicolumn{2}{c|}{SCCE}&
		\multicolumn{2}{c|}{BSSCE}&
		\multicolumn{2}{c}{BSSL}\\
		\hline
		\multicolumn{1}{l|}{Obs}&
		\multicolumn{2}{c}{Mean}&
		\multicolumn{2}{c|}{Median}&
		\multicolumn{1}{c}{Mean}&
		\multicolumn{1}{c|}{Median}&
		\multicolumn{1}{c}{Mean}&
		\multicolumn{1}{c|}{Median}&
		\multicolumn{1}{c}{Mean}&
		\multicolumn{1}{c}{Median}\\
		\hline
		25 & 2.6 & 3.5 & 2.5 & 3.6 & 20.4 & 20.4 & 20.9 & 20.7 & 15.6 & 19.0 \\ 
		30 & 3.0 & 3.7 & 2.9 & 3.7 & 18.4 & 19.9 & 20.9 & 21.2 & 10.3 & 4.1 \\ 
		35 & 3.3 & 3.8 & 3.3 & 3.8 & 15.8 & 18.5 & 19.6 & 19.8 & 7.2 & 4.1 \\ 
		40 & 3.4 & 3.8 & 3.4 & 3.8 & 11.8 & 10.4 & 16.4 & 18.3 & 4.2 & 4.0 \\ 
		45 & 3.6 & 3.8 & 3.6 & 3.8 & 8.1 & 4.1 & 11.0 & 11.3 & 4.1 & 4.0 \\ 
		50 & 3.7 & 3.8 & 3.6 & 3.8 & 7.5 & 4.1 & 7.8 & 4.2 & 4.0 & 4.0 \\ 
		55 & 3.7 & 3.9 & 3.7 & 3.9 & 4.5 & 4.0 & 4.4 & 4.0 & 4.0 & 4.0 \\ 
		60 & 3.8 & 3.9 & 3.8 & 3.9 & 4.2 & 4.0 & 4.0 & 4.0 & 4.0 & 4.0 \\ 
		65 & 3.8 & 3.9 & 3.8 & 3.9 & 4.1 & 4.0 & 4.0 & 4.0 & 4.0 & 4.0 \\ 
		70 & 3.8 & 3.9 & 3.8 & 3.9 & 4.0 & 4.0 & 4.0 & 4.0 & 4.0 & 4.0 \\ 
		75 & 3.8 & 3.9 & 3.8 & 3.9 & 4.0 & 4.0 & 4.0 & 4.0 & 4.0 & 4.0 \\ 
		\hline
	\end{tabular}
	\subcaption{Dispersion of estimated causal effect: values less than 0.05 are replaced with *}
	\begin{tabular}{l|rrrrr|rrr|rrr|rrr}
		\hline
		\multicolumn{1}{l|}{}&
		\multicolumn{5}{c|}{RBCE}&
		\multicolumn{3}{c|}{SCCE}&
		\multicolumn{3}{c|}{BSSCE}&
		\multicolumn{3}{c}{BSSL}\\
		\hline
		\multicolumn{1}{l|}{Obs}&
		\multicolumn{2}{c}{sd}&
		\multicolumn{2}{c}{MSE}&
		\multicolumn{1}{c|}{CI\%}&
		\multicolumn{1}{c}{sd}&
		\multicolumn{1}{c}{MSE}&
		\multicolumn{1}{c|}{CI\%}&
		\multicolumn{1}{c}{sd}&
		\multicolumn{1}{c}{MSE}&
		\multicolumn{1}{c|}{CI\%}&
		\multicolumn{1}{c}{sd}&
		\multicolumn{1}{c}{MSE}&
		\multicolumn{1}{c}{CI\%}
		\\
		\hline
		25 & 0.4 & 0.5 & 0.5 & 2.1 & 100 & 3.1 & 277.4 & 0 & 3.0 & 295.1 & 0 & 9.3 & 217.6 & 20 \\ 
		30 & 0.4 & 0.5 & 0.2 & 1.3 & 100 & 5.9 & 239.9 & 15 & 2.7 & 291.9 & 0 & 8.4 & 107.2 & 60 \\ 
		35 & 0.3 & 0.4 & 0.1 & 0.7 & 100 & 7.2 & 187.0 & 25 & 4.3 & 261.4 & 10 & 6.7 & 53.7 & 80 \\ 
		40 & 0.2 & 0.3 & 0.1 & 0.4 & 100 & 7.8 & 118.7 & 45 & 6.4 & 191.7 & 20 & 0.9 & 0.7 & 95 \\ 
		45 & 0.2 & 0.2 & 0.1 & 0.2 & 100 & 6.0 & 50.7 & 60 & 6.6 & 90.5 & 45 & 0.5 & 0.3 & 95 \\ 
		50 & 0.1 & 0.2 & * & 0.1 & 100 & 5.3 & 39.6 & 65 & 5.5 & 43.0 & 65 & 0.1 & * & 100 \\ 
		55 & 0.1 & 0.1 & * & 0.1 & 100 & 1.6 & 2.6 & 95 & 1.6 & 2.5 & 95 & 0.1 & * & 95 \\ 
		60 & 0.1 & 0.1 & * & 0.1 & 100 & 0.9 & 0.9 & 90 & 0.1 & * & 100 & * & * & 95 \\ 
		65 & 0.1 & 0.1 & * & * & 100 & 0.5 & 0.3 & 95 & 0.1 & * & 95 & * & * & 95 \\ 
		70 & 0.1 & 0.1 & * & * & 100 & * & * & 95 & * & * & 95 & * & * & 95 \\ 
		75 & 0.1 & 0.1 & * & * & 100 & * & * & 95 & * & * & 95 & * & * & 95 \\ 
		\hline
	\end{tabular}
	\subcaption{Accuracy of variable selection: all the values are averaged over 20 replications}
	\begin{tabular}{l|rrr|rr|rr|rr}
		\hline
		\multicolumn{1}{l|}{}&
		\multicolumn{3}{c|}{RBCE}&
		\multicolumn{2}{c|}{SCCE}&
		\multicolumn{2}{c|}{BSSCE}&
		\multicolumn{2}{c}{BSSL}\\
		\hline
		\multicolumn{1}{l|}{Obs}&
		\multicolumn{1}{c}{FP}&
		\multicolumn{1}{c}{FN}&
		\multicolumn{1}{c|}{ID}&
		\multicolumn{1}{c}{FP}&
		\multicolumn{1}{c|}{FN}&
		\multicolumn{1}{c}{FP}&
		\multicolumn{1}{c|}{FN}&
		\multicolumn{1}{c}{FP}&
		\multicolumn{1}{c}{FN}
		\\
		\hline
		25 & 0.7 & 0.2 & 30.4 & 0 & 9.8 & 0 & 10 & 1.4 & 7.1 \\ 
		30 & 0 & 0 & 19.2 & 0 & 8.8 & 0 & 10 & 0 & 3.8 \\ 
		35 & 0 & 0 & 8.1 & 0 & 7.3 & 0 & 9.4 & 0 & 2.0 \\ 
		40 & 0 & 0 & 3.0 & 0 & 5.1 & 0 & 8.2 & 0 & 0.2 \\ 
		45 & 0 & 0 & 0.6 & 0 & 2.9 & 0 & 5.0 & 0 & 0 \\ 
		50 & 0 & 0 & 0.4 & 0 & 2.9 & 0 & 3.0 & 0 & 0 \\ 
		55 & 0 & 0 & 0 & 0 & 0.6 & 0 & 0.4 & 0 & 0 \\ 
		60 & 0 & 0 & 0 & 0 & 0.2 & 0 & 0 & 0 & 0 \\ 
		65 & 0 & 0 & 0 & 0 & 0.2 & 0 & 0 & 0 & 0 \\ 
		70 & 0 & 0 & 0 & 0 & 0 & 0 & 0 & 0 & 0 \\ 
		75 & 0 & 0 & 0 & 0 & 0 & 0 & 0 & 0 & 0 \\ 
		\hline
	\end{tabular}
\end{table}

\paragraph{Results}
\Cref{tab:causal1a} shows the results of estimating the causal effect $\beta_T$
for \added{case 1a}.
For reference, recall the true value is $\beta_T=4$. 
As we perform a sensitivity analysis,
our method gives an interval estimate for the causal effect.
\added{So we present mean, median, sd and MSE for RBCE using two columns where the left columns give the lower bounds
and the right columns give the upper bounds}. We notice that 
\added{as we increase the number of observation our approach provides more precise estimates which shows that our set of priors is able to learn from the data. We also observe that our method tends to under estimate the causal effect. However, our approach does not produce any extreme values} and is more consistent
in terms of estimating the causal effect, \added{especially for fewer number of observations which is not the case for SSCE and BSSCE. For BSSL we notice that median value is close to the true value for fewer number of observations but mean value is higher which shows that for some experiments BSSL tend to produce extreme values. This can also be from the table for dispersion in estimation where BSSL tends to have a high MSE and lower CI\% for fewer number of observations.}

\added{We also provide the performance in variable selection in \cref{tab:causal1a}. We notice that for fewer number of observations
our method tends to give many indeterminate variables but this number gradually decreases as we increase the number of observations. However, our elicitation based approach ensures that we have very few false negative and false positive variables which is not the case for other approaches.}

\begin{table}[ht]
	\tiny
	\centering
	\caption{\added{Comparison of different methods in estimating the causal effect for varying number of observations where some variables are only related to the outcome model.}}
	\label{tab:causal1b}
	\subcaption{Accuracy in estimation of causal effect}
	\begin{tabular}{c|cccc|cc|cc|cc}
		\hline
		\multicolumn{1}{c|}{}&
		\multicolumn{4}{c|}{RBCE}&
		\multicolumn{2}{c|}{SCCE}&
		\multicolumn{2}{c|}{BSSCE}&
		\multicolumn{2}{c}{BSSL}\\
		\hline
		\multicolumn{1}{c|}{Obs}&
		\multicolumn{2}{c}{Mean}&
		\multicolumn{2}{c|}{Median}&
		\multicolumn{1}{c}{Mean}&
		\multicolumn{1}{c|}{Median}&
		\multicolumn{1}{c}{Mean}&
		\multicolumn{1}{c|}{Median}&
		\multicolumn{1}{c}{Mean}&
		\multicolumn{1}{c}{Median}\\
		\hline
		25 & 2.8 & 3.7 & 2.8 & 3.8 & 25.5 & 25.0 & 26.2 & 25.5 & 20.5 & 24.1 \\ 
		30 & 3.2 & 4.1 & 3.1 & 4.1 & 25.4 & 25.1 & 26.4 & 26.2 & 15.8 & 10.8 \\ 
		35 & 3.7 & 4.3 & 3.7 & 4.3 & 22.3 & 25.1 & 26.2 & 26.0 & 10.9 & 4.0 \\ 
		40 & 3.9 & 4.3 & 3.9 & 4.4 & 21.4 & 24.6 & 26.0 & 25.4 & 4.7 & 4.0 \\ 
		45 & 3.9 & 4.2 & 3.9 & 4.2 & 18.7 & 23.9 & 22.5 & 25.2 & 4.0 & 4.0 \\ 
		50 & 3.9 & 4.2 & 3.9 & 4.2 & 9.6 & 4.0 & 13.2 & 6.2 & 4.0 & 4.0 \\ 
		55 & 4.0 & 4.1 & 3.9 & 4.2 & 7.1 & 4.0 & 9.3 & 4.1 & 4.0 & 4.0 \\ 
		60 & 4.0 & 4.1 & 4.0 & 4.1 & 4.0 & 4.0 & 4.0 & 4.0 & 4.0 & 4.0 \\ 
		65 & 4.0 & 4.1 & 4.0 & 4.1 & 4.6 & 4.0 & 4.5 & 4.0 & 4.0 & 4.0 \\ 
		70 & 4.0 & 4.1 & 4.0 & 4.1 & 4.0 & 4.0 & 4.0 & 4.0 & 4.0 & 4.0 \\ 
		75 & 4.0 & 4.1 & 4.0 & 4.1 & 4.0 & 4.0 & 4.0 & 4.0 & 4.0 & 4.0 \\ 
		\hline
	\end{tabular}
	\subcaption{Dispersion of estimated causal effect: values less than 0.05 are replaced with *}
	\begin{tabular}{c|rrrrr|rrr|rrr|rrr}
		\hline
		\multicolumn{1}{c|}{}&
		\multicolumn{5}{c|}{RBCE}&
		\multicolumn{3}{c|}{SCCE}&
		\multicolumn{3}{c|}{BSSCE}&
		\multicolumn{3}{c}{BSSL}\\
		\hline
		\multicolumn{1}{c|}{Obs}&
		\multicolumn{2}{c}{sd}&
		\multicolumn{2}{c}{MSE}&
		\multicolumn{1}{c|}{CI\%}&
		\multicolumn{1}{c}{sd}&
		\multicolumn{1}{c}{MSE}&
		\multicolumn{1}{c|}{CI\%}&
		\multicolumn{1}{c}{sd}&
		\multicolumn{1}{c}{MSE}&
		\multicolumn{1}{c|}{CI\%}&
		\multicolumn{1}{c}{sd}&
		\multicolumn{1}{c}{MSE}&
		\multicolumn{1}{c}{CI\%}
		\\
		\hline
		25 & 0.4 & 0.6 & 0.4 & 1.6 & 100 & 5.6 & 492.2 & 10 & 5.2 & 517.1 & 0 & 10.4 & 375.8 & 10 \\ 
		30 & 0.6 & 0.6 & 0.3 & 0.9 & 100 & 5.8 & 488.8 & 5 & 5.0 & 525.3 & 0 & 12.2 & 280.9 & 50 \\ 
		35 & 0.6 & 0.6 & 0.3 & 0.5 & 100 & 8.5 & 405.3 & 20 & 4.7 & 514.7 & 0 & 9.4 & 130.9 & 60 \\ 
		40 & 0.4 & 0.5 & 0.2 & 0.3 & 100 & 9.6 & 390.6 & 20 & 4.4 & 500.1 & 0 & 3.4 & 11.3 & 95 \\ 
		45 & 0.3 & 0.3 & 0.1 & 0.1 & 100 & 11.0 & 330.5 & 30 & 8.5 & 413.2 & 15 & * & * & 100 \\ 
		50 & 0.3 & 0.3 & 0.1 & 0.1 & 100 & 9.8 & 123.0 & 70 & 11.0 & 199.1 & 50 & * & * & 100 \\ 
		55 & 0.2 & 0.3 & * & 0.1 & 100 & 7.5 & 63.4 & 85 & 9.4 & 110.9 & 75 & * & * & 100 \\ 
		60 & 0.2 & 0.2 & * & 0.1 & 100 & * & * & 100 & * & * & 100 & * & * & 100 \\ 
		65 & 0.1 & 0.2 & * & * & 100 & 2.4 & 6.0 & 95 & 2.1 & 4.3 & 95 & * & * & 100 \\ 
		70 & 0.1 & 0.1 & * & * & 100 & * & * & 100 & * & * & 100 & * & * & 100 \\ 
		75 & 0.1 & 0.1 & * & * & 100 & * & * & 100 & * & * & 100 & * & * & 100 \\ 
		\hline
	\end{tabular}
	\subcaption{Accuracy of variable selection: all the values are averaged over 20 replications}
	\begin{tabular}{c|rrr|rr|rr|rr}
		\hline
		\multicolumn{1}{l|}{}&
		\multicolumn{3}{c|}{RBCE}&
		\multicolumn{2}{c|}{SCCE}&
		\multicolumn{2}{c|}{BSSCE}&
		\multicolumn{2}{c}{BSSL}\\
		\hline
		\multicolumn{1}{c|}{Obs}&
		\multicolumn{1}{c}{FP}&
		\multicolumn{1}{c}{FN}&
		\multicolumn{1}{c|}{ID}&
		\multicolumn{1}{c}{FP}&
		\multicolumn{1}{c|}{FN}&
		\multicolumn{1}{c}{FP}&
		\multicolumn{1}{c|}{FN}&
		\multicolumn{1}{c}{FP}&
		\multicolumn{1}{c}{FN}
		\\
		\hline
		25 & 1.8 & 0.6 & 31.8 & 0 & 14.7 & 0 & 14.9 & 2.0 & 11.6 \\ 
		30 & 0.6 & 0.4 & 27.2 & 0 & 14.8 & 0 & 14.9 & 1.0 & 7.7 \\ 
		35 & 0.2 & 0.5 & 16.9 & 0 & 13.1 & 0 & 15.0 & 0 & 5.2 \\ 
		40 & 0 & 0.2 & 10 & 0 & 12.8 & 0 & 15.0 & 0 & 0.6 \\ 
		45 & 0 & 0 & 2.9 & 0 & 10.2 & 0 & 12.9 & 0 & 0 \\ 
		50 & 0 & 0 & 0.9 & 0 & 3.9 & 0 & 6.6 & 0 & 0 \\ 
		55 & 0 & 0 & 0.4 & 0 & 2.0 & 0 & 3.5 & 0 & 0 \\ 
		60 & 0 & 0 & 0.3 & 0 & 0 & 0 & 0 & 0 & 0 \\ 
		65 & 0 & 0 & 0.3 & 0 & 0.6 & 0 & 0.6 & 0 & 0 \\ 
		70 & 0 & 0 & 0 & 0 & 0 & 0 & 0 & 0 & 0 \\ 
		75 & 0 & 0 & 0 & 0 & 0 & 0 & 0 & 0 & 0 \\ 
		\hline
	\end{tabular}
\end{table}

\added{We present our analysis for case 1b in \cref{tab:causal1b}. Similar to our analyses for case 1a, we notice that as we obtain more observations the imprecision in the estimation reduces. However, unlike case 1a, our approach tends to over estimate the causal effect for higher number of observations. This also shows an overall increasing trend of the estimated causal effect similar to case 1a. We also notice that for case 1b number of indeterminate variables is higher than that of case 1a. This happens as some of the variables are only related to the outcome model. This also contributes to higher number of false negative variables in for other methods. We also observe that similar to the previous case, other methods often produces extreme values for the causal effect increasing the sd and MSE of the estimated causal effect.}

\begin{figure}[h]
	\centering
	\includegraphics[width = 0.9\linewidth]{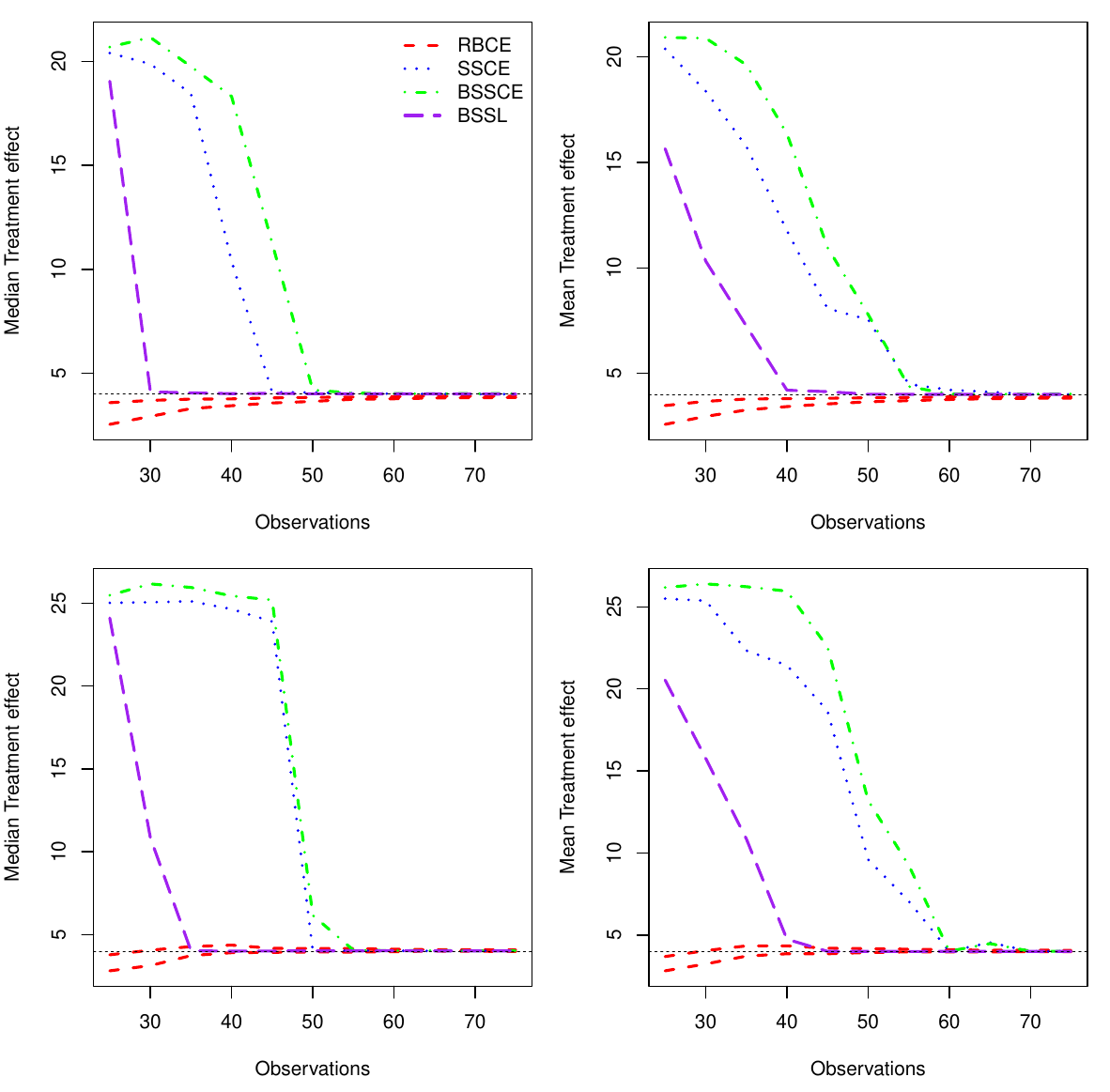}
	\caption{\added{Comparison of different methods in estimating the causal effect for varying number of observations. The top (bottom) row represents case 1a (case 1b). The left(right) images show the average(median) causal effects obtained from 20 replications. }}
	\label{fig:comp:trt}
\end{figure}

\added{We illustrate the estimated causal effect} in \cref{fig:comp:trt} as well. \added{In the figure, the top row
illustrates the case 1a where the left image shows the average value of the estimated
causal effect with respect to observations and the right image
shows the median value. Similarly, the
bottom row represents the same for case 1b. In the figure, RBCE bounds are given by red lines; SSCE estimates by blue lines; BSSCE estimates by green lines; BSSL estimates by purple line; and true value by black lines. In the figure, we can also notice the increase trend of the estimated causal effect as we obtain more observations and also the estimation becomes more precise.}

\begin{figure}[h]
	\centering
	\includegraphics[width = 0.95\linewidth]{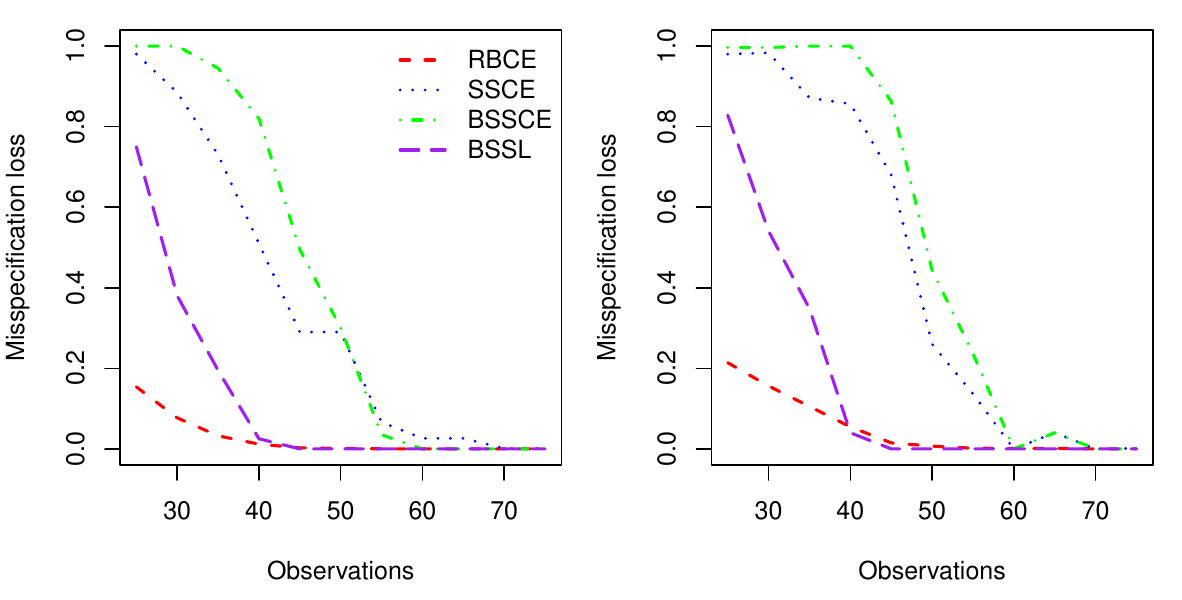}
	\caption{\added{Comparison of different methods in identifying the confounders for varying number of observations. One the left (right) we present case 1a (1b). The red line presents RBCE; blue line presents SSCE; green line presents BSSCE; and purple line
			presents BSSL}}
	\label{fig:comp:loss:obs}
\end{figure}

We also illustrate the performance of variable selection in \cref{fig:comp:loss:obs}.
For variable selection, we use a loss function as described earlier.
Here, we consider
$\ell_1=\ell_2=1$ and $\ell_3= 0.2$,
i.e. we associate a loss of $1$ with false positive and false negative selections,
and a loss of $0.2$ with indeterminate selections. 
Note that we could also choose more 
sophisticated loss functions based on \citep{ZAFFALON20121282}.
We \added{evaluate the misspecification} loss \added{using the equation that we described before}.
From the \added{figure} it can be seen that for \added{case 1}, our method
abstains from identifying some variables for 
$n \le 40$. However,
later on our method gives more precise results in terms of variable
selection. We also notice that the SSCE, BSSCE and BSSL tend to perform poorly
in terms of variable selection. 
\added{However, BSSL performs better than the rest for higher number of observations} 

\begin{table}[ht]
	\tiny
	\centering
	\caption{\added{Comparison of different methods in estimating the causal effect for varying number of predictors where all the active variables are confounders.}}
	\label{tab:causal2a}
	\subcaption{Accuracy in estimation of causal effect}
	\begin{tabular}{c|cccc|cc|cc|cc}
		\hline
		\multicolumn{1}{l|}{}&
		\multicolumn{4}{c|}{RBCE}&
		\multicolumn{2}{c|}{SCCE}&
		\multicolumn{2}{c|}{BSSCE}&
		\multicolumn{2}{c}{BSSL}\\
		\hline
		\multicolumn{1}{c|}{Pred}&
		\multicolumn{2}{c}{Mean}&
		\multicolumn{2}{c|}{Median}&
		\multicolumn{1}{c}{Mean}&
		\multicolumn{1}{c|}{Median}&
		\multicolumn{1}{c}{Mean}&
		\multicolumn{1}{c|}{Median}&
		\multicolumn{1}{c}{Mean}&
		\multicolumn{1}{c}{Median}\\
		\hline
		1 & 2 & 3 & 4 & 5 & 6 & 7 & 8 & 9 & 10 & 11 \\ 
		\hline
		25 & 3.6 & 3.8 & 3.7 & 3.8 & 9.5 & 4.6 & 11.8 & 14.0 & 4.0 & 4.0 \\ 
		30 & 3.6 & 3.8 & 3.6 & 3.8 & 8.8 & 4.1 & 9.7 & 4.7 & 4.5 & 4.0 \\ 
		35 & 3.6 & 3.8 & 3.6 & 3.8 & 12.8 & 17.4 & 12.8 & 14.9 & 4.9 & 4.0 \\ 
		40 & 3.5 & 3.8 & 3.5 & 3.8 & 12.3 & 15.2 & 14.9 & 17.6 & 4.0 & 4.0 \\ 
		45 & 3.5 & 3.8 & 3.4 & 3.8 & 10.7 & 4.1 & 16.6 & 18.3 & 4.0 & 4.0 \\ 
		50 & 3.4 & 3.8 & 3.5 & 3.8 & 11.8 & 10.4 & 16.4 & 18.3 & 4.2 & 4.0 \\ 
		55 & 3.4 & 3.8 & 3.4 & 3.8 & 12.4 & 15.6 & 16.6 & 18.8 & 4.5 & 4.0 \\ 
		60 & 3.3 & 3.8 & 3.3 & 3.8 & 14.2 & 18.7 & 17.3 & 19.1 & 4.1 & 4.0 \\ 
		65 & 3.2 & 3.8 & 3.3 & 3.8 & 12.6 & 15.5 & 19.4 & 19.2 & 5.0 & 4.0 \\ 
		70 & 3.1 & 3.8 & 3.2 & 3.8 & 11.1 & 4.1 & 16.7 & 19.1 & 4.0 & 4.0 \\ 
		75 & 3.1 & 3.8 & 3.1 & 3.8 & 11.7 & 9.3 & 18.4 & 19.2 & 4.5 & 4.0 \\ 
		\hline
	\end{tabular}
	\subcaption{Dispersion of estimated causal effect: values less than 0.05 are replaced with *}
	\begin{tabular}{c|rrrrr|rrr|rrr|rrr}
		\hline
		\multicolumn{1}{l|}{}&
		\multicolumn{5}{c|}{RBCE}&
		\multicolumn{3}{c|}{SCCE}&
		\multicolumn{3}{c|}{BSSCE}&
		\multicolumn{3}{c}{BSSL}\\
		\hline
		\multicolumn{1}{c|}{Pred}&
		\multicolumn{2}{c}{sd}&
		\multicolumn{2}{c}{MSE}&
		\multicolumn{1}{c|}{CI\%}&
		\multicolumn{1}{c}{sd}&
		\multicolumn{1}{c}{MSE}&
		\multicolumn{1}{c|}{CI\%}&
		\multicolumn{1}{c}{sd}&
		\multicolumn{1}{c}{MSE}&
		\multicolumn{1}{c|}{CI\%}&
		\multicolumn{1}{c}{sd}&
		\multicolumn{1}{c}{MSE}&
		\multicolumn{1}{c}{CI\%}
		\\
		\hline
		25 & 0.2 & 0.2 & 0.1 & 0.2 & 100 & 6.6 & 72.1 & 50 & 7.1 & 109.3 & 45 & 0.1 & * & 100 \\ 
		30 & 0.2 & 0.2 & 0.1 & 0.2 & 100 & 6.6 & 64.4 & 65 & 6.8 & 76.7 & 60 & 2.0 & 3.9 & 95 \\ 
		35 & 0.2 & 0.2 & 0.1 & 0.2 & 100 & 7.4 & 129.2 & 35 & 7.4 & 130.1 & 40 & 4.1 & 16.6 & 90 \\ 
		40 & 0.2 & 0.2 & 0.1 & 0.2 & 100 & 7.9 & 129.0 & 45 & 6.7 & 162.0 & 30 & 0.1 & * & 95 \\ 
		45 & 0.2 & 0.2 & 0.1 & 0.3 & 100 & 7.8 & 102.5 & 55 & 5.7 & 190.4 & 20 & 0.1 & * & 100 \\ 
		50 & 0.2 & 0.3 & 0.1 & 0.4 & 100 & 7.8 & 118.7 & 45 & 6.4 & 191.7 & 20 & 0.9 & 0.7 & 95 \\ 
		55 & 0.2 & 0.3 & 0.1 & 0.5 & 100 & 8.0 & 132.4 & 45 & 6.8 & 201.9 & 25 & 2.3 & 5.2 & 95 \\ 
		60 & 0.2 & 0.3 & 0.1 & 0.6 & 100 & 8.1 & 165.8 & 30 & 6.2 & 214.7 & 15 & 0.6 & 0.4 & 100 \\ 
		65 & 0.2 & 0.4 & 0.1 & 0.8 & 100 & 8.3 & 139.2 & 40 & 2.8 & 244.0 & 0 & 4.4 & 19.1 & 95 \\ 
		70 & 0.2 & 0.4 & 0.1 & 0.9 & 100 & 8.2 & 114.9 & 55 & 6.6 & 202.1 & 25 & 0.1 & * & 95 \\ 
		75 & 0.2 & 0.4 & 0.1 & 1.0 & 100 & 8.0 & 120.2 & 50 & 4.8 & 228.7 & 10 & 3.5 & 11.9 & 85 \\ 
		\hline
	\end{tabular}
	\subcaption{Accuracy of variable selection: all the values are averaged over 20 replications}
	\begin{tabular}{c|rrr|rr|rr|rr}
		\hline
		\multicolumn{1}{l|}{}&
		\multicolumn{3}{c|}{RBCE}&
		\multicolumn{2}{c|}{SCCE}&
		\multicolumn{2}{c|}{BSSCE}&
		\multicolumn{2}{c}{BSSL}\\
		\hline
		\multicolumn{1}{c|}{Pred}&
		\multicolumn{1}{c}{FP}&
		\multicolumn{1}{c}{FN}&
		\multicolumn{1}{c|}{ID}&
		\multicolumn{1}{c}{FP}&
		\multicolumn{1}{c|}{FN}&
		\multicolumn{1}{c}{FP}&
		\multicolumn{1}{c|}{FN}&
		\multicolumn{1}{c}{FP}&
		\multicolumn{1}{c}{FN}
		\\
		\hline
		25 & 0 & 0 & 5.2 & 0 & 4.0 & 0 & 5.5 & 0 & 0 \\ 
		30 & 0 & 0 & 2.4 & 0 & 3.5 & 0 & 4.1 & 0 & 0.4 \\ 
		35 & 0 & 0 & 2.0 & 0 & 6.0 & 0 & 6.0 & 0 & 0.5 \\ 
		40 & 0 & 0 & 2.1 & 0 & 5.4 & 0 & 7.3 & 0 & 0 \\ 
		45 & 0 & 0 & 2.6 & 0 & 4.4 & 0 & 8.4 & 0 & 0 \\ 
		50 & 0 & 0 & 3.0 & 0 & 5.1 & 0 & 8.2 & 0 & 0.2 \\ 
		55 & 0 & 0 & 3.8 & 0 & 5.4 & 0 & 7.9 & 0 & 0.4 \\ 
		60 & 0 & 0 & 5.0 & 0 & 6.4 & 0 & 8.4 & 0 & 0 \\ 
		65 & 0 & 0 & 4.4 & 0 & 5.4 & 0 & 9.9 & 0 & 0.5 \\ 
		70 & 0 & 0 & 4.8 & 0 & 4.4 & 0 & 8.0 & 0 & 0 \\ 
		75 & 0 & 0 & 6.2 & 0 & 4.9 & 0 & 9.2 & 2.4 & 0.6 \\ 
		\hline
	\end{tabular}
\end{table}

We show the result of our analyses \added{case 2a} in \cref{tab:causal2a}. Similar to
our analyses with increasing number of observations, we notice that our method 
is overall in agreement with
\added{BSSL}. However, \added{similar to case 1a} our method tends to
underestimate the treatment effect (approximately 5\%) for \added{case 2b}. \added{We also notice that the imprecision in estimation increases as we increase the number of predictors. This happens as observation per predictor reduces. We also notice that BSSL outperforms RBCE in terms of median value of estimated causal effect over 20 replications. However, in very few cases BSSL provides extreme values which can be understood from mean and CI\% as well as MSE. Moreover, for 75 predictors BSSL gives higher number of false positives which is not the case for RBCE. Unlike the case 1a and 1b, SSCE and BSSCE performs poorly for every value of predictors.}

\begin{table}[h]
	\tiny
	\centering
	\caption{\added{Comparison of different methods in estimating the causal effect for varying number of predictors where some variables are only to the outcome model.}}
	\label{tab:causal2b}
	\subcaption{Accuracy in estimation of causal effect}
	\begin{tabular}{c|cccc|cc|cc|cc}
		\hline
		\multicolumn{1}{c|}{}&
		\multicolumn{4}{c|}{RBCE}&
		\multicolumn{2}{c|}{SCCE}&
		\multicolumn{2}{c|}{BSSCE}&
		\multicolumn{2}{c}{BSSL}\\
		\hline
		\multicolumn{1}{c|}{Pred}&
		\multicolumn{2}{c}{Mean}&
		\multicolumn{2}{c|}{Median}&
		\multicolumn{1}{c}{Mean}&
		\multicolumn{1}{c|}{Median}&
		\multicolumn{1}{c}{Mean}&
		\multicolumn{1}{c|}{Median}&
		\multicolumn{1}{c}{Mean}&
		\multicolumn{1}{c}{Median}\\
		\hline
		1 & 2 & 3 & 4 & 5 & 6 & 7 & 8 & 9 & 10 & 11 \\ 
		\hline
		25 & 3.8 & 4.1 & 3.8 & 4.1 & 19.0 & 22.3 & 24.0 & 25.3 & 4.0 & 4.0 \\ 
		30 & 3.9 & 4.2 & 3.9 & 4.1 & 20.6 & 24.6 & 20.9 & 24.5 & 7.4 & 4.0 \\ 
		35 & 3.9 & 4.3 & 3.9 & 4.2 & 21.9 & 24.7 & 23.7 & 24.8 & 7.3 & 4.0 \\ 
		40 & 3.9 & 4.3 & 3.9 & 4.3 & 20.9 & 24.4 & 24.7 & 25.5 & 5.1 & 4.0 \\ 
		45 & 3.8 & 4.3 & 3.9 & 4.3 & 20.2 & 23.9 & 26.1 & 25.4 & 4.8 & 4.0 \\ 
		50 & 3.9 & 4.3 & 3.9 & 4.3 & 21.4 & 24.6 & 26.0 & 25.4 & 4.7 & 4.0 \\ 
		55 & 3.8 & 4.4 & 3.9 & 4.4 & 21.8 & 24.8 & 26.0 & 25.1 & 9.5 & 4.0 \\ 
		60 & 3.8 & 4.4 & 3.9 & 4.4 & 16.6 & 19.7 & 25.1 & 25.5 & 8.4 & 4.0 \\ 
		65 & 3.7 & 4.4 & 3.8 & 4.4 & 18.7 & 22.8 & 26.2 & 25.6 & 5.4 & 4.0 \\ 
		70 & 3.7 & 4.5 & 3.8 & 4.4 & 20.4 & 23.9 & 25.5 & 25.2 & 8.5 & 4.0 \\ 
		75 & 3.7 & 4.5 & 3.7 & 4.4 & 17.7 & 22.8 & 26.0 & 25.2 & 5.0 & 4.0 \\ 
		\hline
	\end{tabular}
	\subcaption{Dispersion of estimated causal effect: values less than 0.05 are replaced with *}
	\begin{tabular}{c|rrrrr|rrr|rrr|rrr}
		\hline
		\multicolumn{1}{c|}{}&
		\multicolumn{5}{c|}{RBCE}&
		\multicolumn{3}{c|}{SCCE}&
		\multicolumn{3}{c|}{BSSCE}&
		\multicolumn{3}{c}{BSSL}\\
		\hline
		\multicolumn{1}{c|}{Pred}&
		\multicolumn{2}{c}{sd}&
		\multicolumn{2}{c}{MSE}&
		\multicolumn{1}{c|}{CI\%}&
		\multicolumn{1}{c}{sd}&
		\multicolumn{1}{c}{MSE}&
		\multicolumn{1}{c|}{CI\%}&
		\multicolumn{1}{c}{sd}&
		\multicolumn{1}{c}{MSE}&
		\multicolumn{1}{c|}{CI\%}&
		\multicolumn{1}{c}{sd}&
		\multicolumn{1}{c}{MSE}&
		\multicolumn{1}{c}{CI\%}
		\\
		\hline
		25 & 0.2 & 0.3 & 0.1 & 0.1 & 100 & 11.2 & 343.7 & 35 & 8.0 & 459.5 & 10 & 0.1 & * & 100 \\ 
		30 & 0.3 & 0.4 & 0.1 & 0.1 & 100 & 10.4 & 377.0 & 25 & 10.8 & 397.7 & 25 & 8.3 & 77.9 & 85 \\ 
		35 & 0.3 & 0.4 & 0.1 & 0.2 & 100 & 9.9 & 412.9 & 20 & 7.4 & 438.6 & 10 & 8.0 & 71.7 & 85 \\ 
		40 & 0.4 & 0.4 & 0.1 & 0.3 & 100 & 9.3 & 366.8 & 15 & 7.0 & 475.8 & 10 & 4.8 & 23.1 & 95 \\ 
		45 & 0.4 & 0.4 & 0.2 & 0.3 & 100 & 10.4 & 365.5 & 25 & 4.4 & 504.8 & 0 & 3.4 & 11.3 & 95 \\ 
		50 & 0.4 & 0.5 & 0.2 & 0.3 & 100 & 9.6 & 390.6 & 20 & 4.4 & 500.1 & 0 & 3.4 & 11.3 & 95 \\ 
		55 & 0.4 & 0.5 & 0.2 & 0.3 & 100 & 9.9 & 410.6 & 20 & 4.5 & 503.9 & 0 & 9.9 & 123.1 & 75 \\ 
		60 & 0.5 & 0.5 & 0.2 & 0.4 & 100 & 11.7 & 288.5 & 40 & 6.4 & 484.0 & 5 & 9.3 & 102.4 & 80 \\ 
		65 & 0.4 & 0.5 & 0.2 & 0.4 & 100 & 11.2 & 334.4 & 30 & 4.5 & 510.0 & 0 & 6.4 & 40.5 & 95 \\ 
		70 & 0.5 & 0.5 & 0.2 & 0.5 & 100 & 11.1 & 385.3 & 25 & 5.2 & 488.8 & 5 & 9.3 & 102.1 & 80 \\ 
		75 & 0.5 & 0.5 & 0.2 & 0.5 & 100 & 11.8 & 318.1 & 40 & 4.5 & 502.1 & 0 & 4.0 & 16.3 & 95 \\ 
		\hline
	\end{tabular}
	\subcaption{Accuracy of variable selection: all the values are averaged over 20 replications}
	\begin{tabular}{c|rrr|rr|rr|rr}
		\hline
		\multicolumn{1}{c|}{}&
		\multicolumn{3}{c|}{RBCE}&
		\multicolumn{2}{c|}{SCCE}&
		\multicolumn{2}{c|}{BSSCE}&
		\multicolumn{2}{c}{BSSL}\\
		\hline
		\multicolumn{1}{c|}{Pred}&
		\multicolumn{1}{c}{FP}&
		\multicolumn{1}{c}{FN}&
		\multicolumn{1}{c|}{ID}&
		\multicolumn{1}{c}{FP}&
		\multicolumn{1}{c|}{FN}&
		\multicolumn{1}{c}{FP}&
		\multicolumn{1}{c|}{FN}&
		\multicolumn{1}{c}{FP}&
		\multicolumn{1}{c}{FN}
		\\
		\hline
		25 & 0 & 0 & 7.8 & 0 & 10.2 & 0 & 13.5 & 0 & 0 \\ 
		30 & 0 & 0 & 8.2 & 0 & 11.6 & 0 & 11.2 & 0 & 2.0 \\ 
		35 & 0 & 0 & 9.2 & 0 & 12.6 & 0 & 14.0 & 0 & 2.2 \\ 
		40 & 0 & 0 & 10.1 & 0 & 11.9 & 0 & 13.8 & 0 & 0.8 \\ 
		45 & 0 & 0.1 & 10.8 & 0 & 11.8 & 0 & 15.0 & 0 & 0.6 \\ 
		50 & 0 & 0.2 & 10.1 & 0 & 12.8 & 0 & 15.0 & 0 & 0.6 \\ 
		55 & 0 & 0.3 & 8.9 & 0 & 12.0 & 0 & 15.0 & 0 & 3.5 \\ 
		60 & 0 & 0.4 & 11.1 & 0 & 9.2 & 0 & 14.4 & 0.5 & 2.9 \\ 
		65 & 0 & 0.8 & 11.9 & 0 & 10.6 & 0 & 15.0 & 0.4 & 0.8 \\ 
		70 & 0 & 0.8 & 12.8 & 0 & 10.9 & 0 & 14.8 & 0 & 2.8 \\ 
		75 & 0 & 1.0 & 12.3 & 0.1 & 9.6 & 0 & 15.0 & 0 & 0.9 \\ 
		\hline
	\end{tabular}
\end{table}

\added{The result for case 2b is presented in \cref{tab:causal2b}. We notice that for this case the true causal effect is always contained within the estimated bounds unlike the previous cases. For this case, the imprecision in the estimated causal effect increases with respect to predictors similar to case 2a as the observation per predictor reduces. We also see that SSCE and BSSCE performs poorly similar to case 2a and BSSL is mostly consistent in estimation but produces extreme values for some experiments giving a significant differences between mean and median of the estimated causal treatments.}

\begin{figure}[h]
	\centering
	\includegraphics[width = 0.85\linewidth]{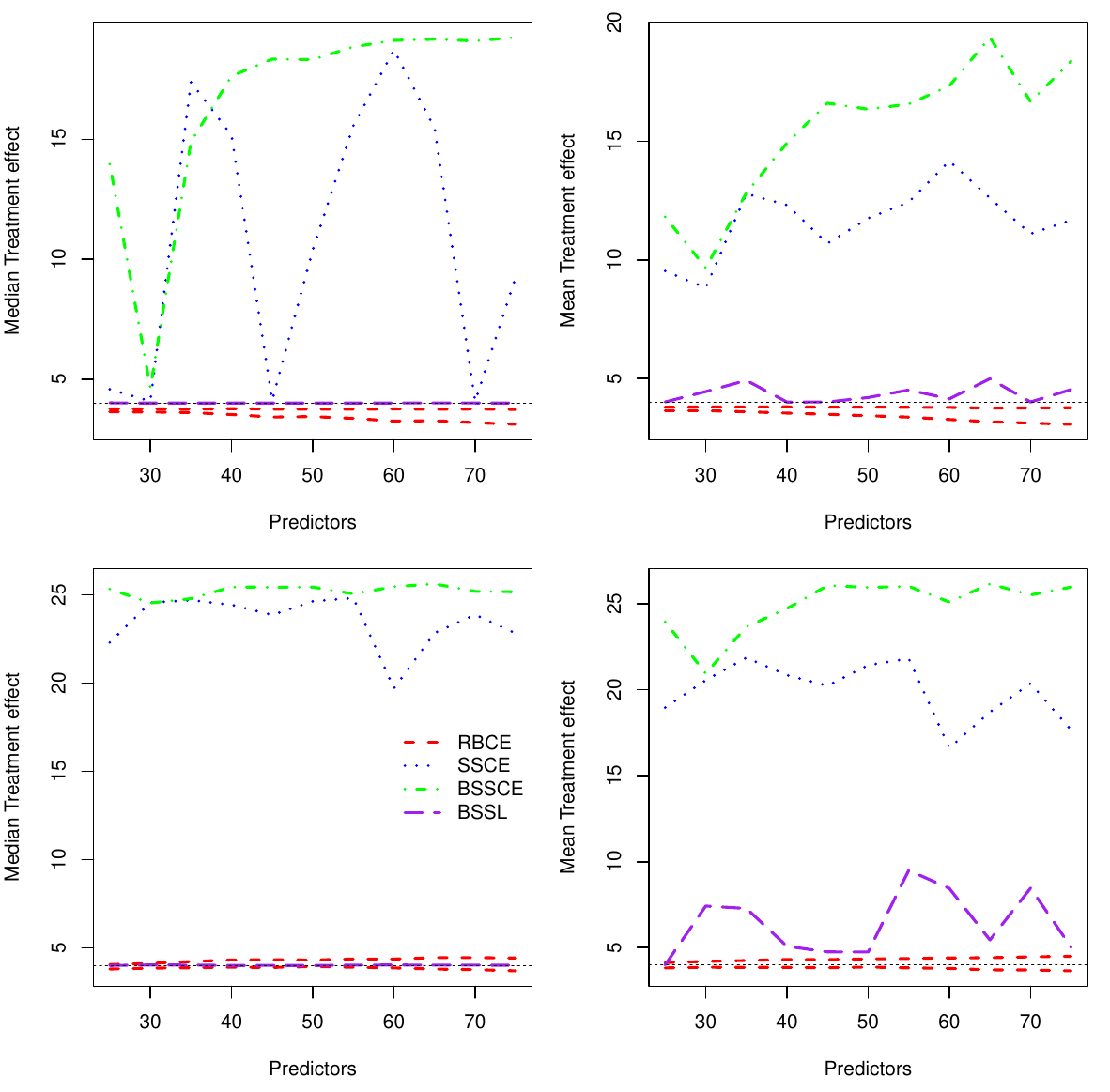}
	\caption{\added{Comparison of different methods in estimating the causal effect for varying number of predictors. The top (bottom) row represents case 2a (case 2b). The left(right) images show the average(median) causal effects obtained from 20 replications. }}
	\label{fig:comp:trt:pred}
\end{figure}

\begin{figure}[h]
	\centering
	\includegraphics[width = 0.9\linewidth]{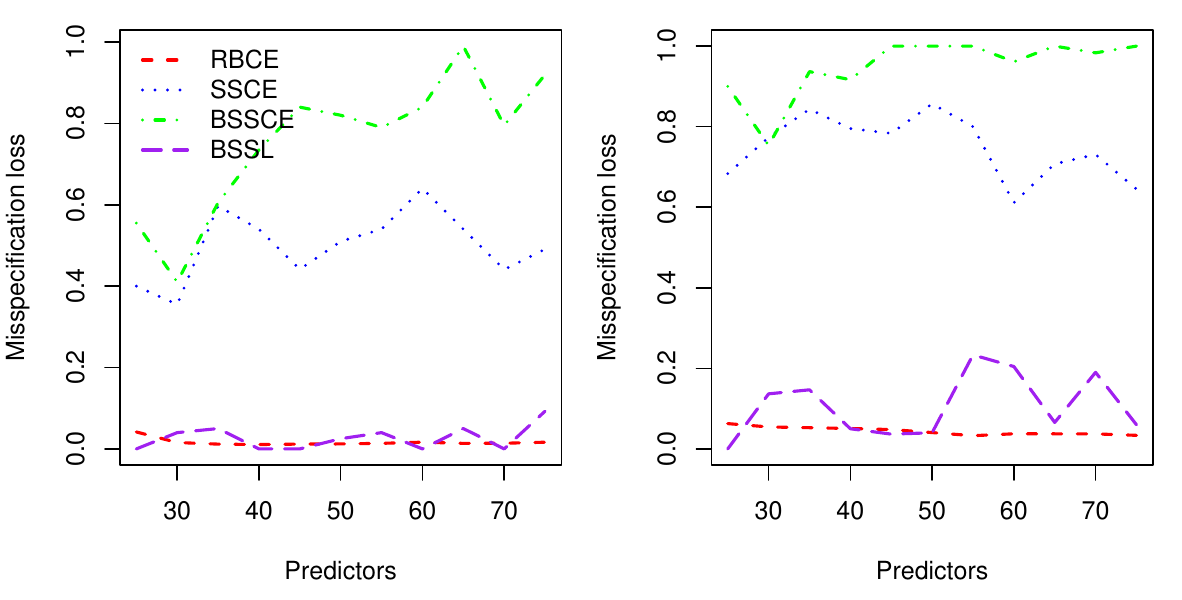}
	\caption{\added{Comparison of different methods in identifying the confounders for varying number of predictors. One the left (right) we present case 1a (1b). The red line presents RBCE; blue line presents SSCE; green line presents BSSCE; and purple line
			presents BSSL}}
	\label{fig:comp:loss:pred}
\end{figure}

\added{We also show the causal effect estimation and performance in variable selection in \cref{fig:comp:trt:pred,fig:comp:loss:pred}. From \cref{fig:comp:trt:pred} we can see the increase in imprecision as we increase then number of predictors. We also see that RBCE performs more consistently than other methods in terms of estimating the causal effect. We also notice that BSSL outperforms RBCE in terms of estimating the causal effect and performs at per in terms of variable selection for case 2a. However, for case 2b, BSSL appears to be less consistent in terms of variable selection. From these two figures we can also see that SSCE and BSSCE performs poorly as we increase the number of predictors and is particularly unstable for case 2a.}

% latex table generated in R 4.3.2 by xtable 1.8-4 package
% Mon Mar 25 10:08:22 2024

\begin{table}[h]
	\caption{\added{Effect of elicitation of the inclusion probability of the variables where FP stands for false positive, FN stands for false negative and ID stands for indeterminate. }}
	\label{tab:comp:elict}\centering
	\begin{tabular}{c|rrrrr|rrrrr}
		\hline
		\multicolumn{1}{c|}{$[\underline{c},\overline{c}]$}&
		\multicolumn{5}{c|}{$[0.15,0.35]$}&
		\multicolumn{5}{c}{$[0.2,0.4]$}\\
		\hline
		\multicolumn{1}{c|}{Pred}&
		\multicolumn{2}{c}{Mean} &
		\multicolumn{1}{c}{FP} &
		\multicolumn{1}{c}{FN} &
		\multicolumn{1}{c|}{IDR} &
		\multicolumn{2}{c}{Mean} &
		\multicolumn{1}{c}{FP} &
		\multicolumn{1}{c}{FN} &
		\multicolumn{1}{c}{IDR}\\
		\hline
		25 & 3.8 & 4.1 & 0 & 0 & 7.8 & 3.9 & 4.2 & 0 & 0 & 4.9 \\ 
		30 & 3.9 & 4.2 & 0 & 0 & 8.2 & 3.9 & 4.3 & 0 & 0 & 4.2 \\ 
		35 & 3.9 & 4.3 & 0 & 0 & 9.2 & 4.0 & 4.3 & 0 & 0.2 & 3.5 \\ 
		40 & 3.9 & 4.3 & 0 & 0 & 10.1 & 4.0 & 4.4 & 0 & 0.3 & 3.4 \\ 
		45 & 3.8 & 4.3 & 0 & 0.1 & 10.8 & 4.0 & 4.4 & 0 & 0.4 & 3.5 \\ 
		50 & 3.9 & 4.3 & 0 & 0.2 & 10.1 & 4.0 & 4.4 & 0 & 0.6 & 3.0 \\ 
		55 & 3.8 & 4.4 & 0 & 0.3 & 8.9 & 4.0 & 4.4 & 0 & 0.6 & 3.0 \\ 
		60 & 3.8 & 4.4 & 0 & 0.4 & 11.1 & 4.0 & 4.5 & 0 & 0.8 & 3.4 \\ 
		65 & 3.7 & 4.4 & 0 & 0.8 & 11.9 & 4.0 & 4.5 & 0 & 1.0 & 3.8 \\ 
		70 & 3.7 & 4.5 & 0 & 0.8 & 12.8 & 4.0 & 4.5 & 0 & 1.3 & 3.0 \\ 
		75 & 3.7 & 4.5 & 0 & 1.0 & 12.3 & 4.0 & 4.5 & 0 & 1.4 & 3.1 \\ 
		\hline
	\end{tabular}
\end{table}

\paragraph{\added{Importance of prior elicitation}}\label{sec:importance:elicit}
\added{Our method relies on expert elicitation and prior sensitivity analysis. So we also explore the effect of prior elicitation in identifying the active variables in our model. As mentioned earlier,
	we consider $c$ is expected to be lying in the interval $[0.15,0.35]$
	based on \cref{tab:corr}. However, we might want to choose a different
	value for $c$. To compare the effect of having different value for $c$, we use the case 2b and set $c\in[0.2,0.4]$. That is we set a higher threshold for the correlation so that $\overline{k}$ becomes smaller and hence the prior expectation of the inclusion probability. We show our results in \cref{tab:comp:elict}. In the left hand side we elicit the expected number of active variables by setting the marginal correlation threshold $c\in[0.15,0.35]$ and the right we set $c\in[0.2,0.4]$. On the left hand side, we see our method tends to give higher number of indeterminate variables for fewer observations than the right hand side. As we increase the predictors the higher threshold of marginal correlation plays an important role and we see more cases of false negative variables on the right hand side. This also results to over estimation of the causal effect as many true active variables are shrinked towards zero. As a result the lower bound of the averaged causal effect is more than four on the right hand side.}

The analyses and simulation studies can be investigated using the code from \url{https://github.com/tathagatabasu/Causal-Inference}.

\section{Conclusion}\label{sec:conc}

Causal effect estimation is an important tool in statistical learning.
Especially in risk-sensitive situations, such as medicine, it needs to
be performed \added{with the utmost care} as in many cases poor estimation can have severe adverse consequences.
In this paper, we tackle this issue by proposing a robust Bayesian analysis of the causal 
effect estimation problem for high dimensional data. Our 
framework is focused on the effect of prior elicitation on
predictor selection
as well as causal effect estimation. We consider a spike and slab type
prior for predictor selection and discuss the possible sources of uncertainty that
need to be tackled carefully. We were particularly focused on the uncertainty associated
with prior selection probabilities for which we consider a set of beta priors to perform
sensitivity analysis. We showed that the sensitivity analysis on the prior selection probability
gives us a robust predictor selection scheme. In this way, we can abstain from selecting
a predictor when the available data is not sufficient. We also propose a more relaxed
utility based framework, where we associate a loss for abstaining which can be interpreted 
as the cost of further data collection. We illustrate our method with synthetic dataset
and compare with other state of the art Bayesian methods. \added{We could see that our elicitation based approach helps to have a more consistent causal effect estimation for very limited number of observations and avoids producing extreme values for the causal effect. Moreover, we also notice correct elicitation of the inclusion probability plays a crucial role in identifying the active variables and therefore can be extremely useful in cases where we need to design a treatment guideline with multiple bio-markers. }

Currently, the paper proposes a robust Bayesian approach for causal effect estimation where
we rely on sampling strategies to obtain the posterior bounds as well as performing 
variable selection.
\added{%
A weakness of our approach is simulation efficiency,
as we resorted to brute force optimisation.
However, there is ample opportunity to improve computational aspects.%
}
\added{In the future,} it will be interesting to derive inner approximation bounds
for the posterior estimates to reduce the computational cost,
or to find better ways than brute force optimisation, such as for instance iterative importance sampling \citep{cruz22_importance}.

\added{To compare the different methods, we rely on simple loss functions
associated with the predictor selection. However,
loss functions could be used for a generalised
decision theoretic framework as well.
For instance,
the selection problem itself could be formulated as a decision problem,
potentially leading to different selection thresholds or even selection systems that are directly based on a loss function.
Additionally, we could formulate the 
problem of whether or not to treat a subject as a decision support problem
based on predictor selection.%
}

\added{%
Another topic of interest pertinent to medical diagnosis is missing data.
It has been shown that using bounded probability is particularly suitable
for dealing with instances where data cannot be assumed missing at random
\citep{decoomanzaffalon2004}.
Incorporating robustness against missing data could lead to an interesting
extension of the model in this paper.%
}

We also notice with our simulation studies that our
method tends to underestimate the causal effect
when \added{only confounders are present in the}
model. This suggests that we might want
to use a correction formula for the causal effect.
Moreover, in future, we would
like to investigate different elicitation strategies for different prior parameters and their importance in causal effect estimation.

In general, we noticed that our method
is in good agreement with other methods with an added level of robustness.
This shows that our method has good potential for real-life problems,
and we intend to apply it on a real dataset in future work.

\section*{Acknowledgements}

We sincerely thank Jochen Einbeck for his contributions to an earlier version of this paper.
We also thank all reviewers for their kind and constructive comments which helped improve the paper.

\bibliographystyle{elsarticle-num-names} 
\bibliography{basu_causal}

\end{document}